\documentclass[11pt]{article}
\usepackage[margin=1in]{geometry}
\usepackage{authblk}
\usepackage{times}
\usepackage[noadjust]{cite}
\usepackage[normalem]{ulem}
\usepackage[english]{babel}
\usepackage{blindtext}
\newcommand{\nnumber}[1]{\textcolor{black}{#1}}

\usepackage{lmodern}
\usepackage{booktabs}
\usepackage{url}
\usepackage{graphicx}
\usepackage{tikz}
\usepackage{amsmath}
\usepackage{epsfig,endnotes, graphicx,booktabs, multirow, array}
\usepackage{xspace}
\usepackage{colortbl}

\usepackage{tabularx}
\usepackage[utf8]{inputenc}
\usepackage{regexpatch}
\usepackage[font=footnotesize]{caption} % abovecaptionskip=1pt
\makeatletter
\newcommand{\linebreakand}{%
  \end{@IEEEauthorhalign}
  \hfill\mbox{}\par
  \mbox{}\hfill\begin{@IEEEauthorhalign}
}
\makeatother

\usepackage{algorithm}
\usepackage[noend]{algpseudocode}

\usepackage{xspace}
\usepackage{hyperref}
\usepackage[textsize=small,backgroundcolor=orange]{todonotes}
\usepackage{booktabs}
\usepackage{graphicx}
\usepackage{enumitem}
\usepackage{multirow,tabularx}

\definecolor{aliceblue}{rgb}{0.94, 0.97, 1.0}

\definecolor{aliceblue}{rgb}{0.94, 0.97, 1.0}
\newcommand{\red}[1]{\textcolor{red}{#1}}

\usepackage{tcolorbox}
\newtcolorbox{mybox}[1]{colback=aliceblue,colframe=black,fonttitle=\bfseries,title=#1}

\hypersetup{
    colorlinks=true,
    linkcolor=blue,
    filecolor=magenta,      
    urlcolor=cyan
    }
\RequirePackage[capitalise]{cleveref}
\usepackage{cleveref} 
\usepackage{cite}
\usepackage{url}                % allow \url in bibtex for clickable links
\usepackage{xcolor}             % color definitions, to be use for...
\hypersetup{                    % ...like so
  colorlinks,
  linkcolor={green!80!black},
  citecolor={red!70!black},
  urlcolor={blue!70!black}
}

\usepackage{mwe}
\usepackage{tikz}
\usetikzlibrary{arrows}
\usepackage{verbatim}

\usepackage{booktabs}
\usepackage{dirtree}
\usepackage{wrapfig}
\usepackage{caption}

\title{\Large \bf The Overview of Privacy Labels and their Compatibility with Privacy Policies}

\renewcommand*{\Affilfont}{\normalsize\normalfont}
\newsavebox\affbox

\usepackage{enumitem}
\usepackage{csquotes}
\usepackage{subfig}
\usepackage{tabularx}
\usepackage{amsmath}
\usepackage{makecell}
\usepackage{multirow}
\definecolor{aliceblue}{rgb}{0.94, 0.97, 1.0}
\usepackage{tikz}
\usepackage{amsmath}
\usepackage{xspace}
\usepackage{colortbl}
\usepackage{authblk}
\usepackage{tabularx}
\usepackage[title]{appendix}
\usepackage[utf8]{inputenc}
\begin{document}
\newcommand*\samethanks[1][\value{footnote}]{\footnotemark[#1]}
\author{Rishabh Khandelwal\thanks{Equal Contribution}}
\author{Asmit Nayak\samethanks}
\author{Paul Chung}
\author{Kassem Fawaz}
\affil[]{%
  \savebox\affbox{\Affilfont Department of Chemical Engineering, University of AAAAA BBBBBB, CCCCC road,}%
  \parbox[t]{\wd\affbox}{\protect\centering} University of Wisconsin -- Madison} 
\date{}

\maketitle

\section{Introduction}
\label{sec:intro}
Privacy policies have traditionally been the primary means of conveying the privacy practices of a service to users. However, studies have shown that privacy policies are often ineffective due to readability and reachability issues, as users tend to avoid reading them due to their length and vagueness ~\cite{cate2010limits, gluck2016short}. Introduced by Kelly et al~\cite{kelley_labels}, the concept of privacy labels has gained traction in the tech industry, with Google introducing Data Safety Sections (DSS) and Apple introducing Apple Privacy Labels (APL) for all new and updated apps on the App Store.

% Prior research has examined the reactiveness of developers in implementing Apple Privacy Labels and analyzed the data collection practices of apps according to these labels ~\cite{balash2022longitudinal, li2022understandingios}, however, they have not provided a comprehensive analysis of the current status of privacy labels for the play store. 

Researchers have shown the benefit of privacy labels for users, making privacy practices more accessible~\cite{zhang2022usable}. However, prior work has also shown that inaccurate labels can exist due to the developer's knowledge gaps or resource limitations~\cite{li2022understanding}. 
Incorrect privacy labels can cause confusion and harm users by creating a false sense of security. Furthermore, inaccurate privacy labels can mislead users into downloading and using insecure apps, increasing their privacy risks. Therefore, it is crucial to investigate the accuracy and compliance of privacy labels in real-world scenarios, in order to determine how well they align with the actual data practices of apps.

Xiao et al~\cite{xiao2022lalaine} proposed a methodology to check for consistency of privacy labels by comparing practices in labels with privacy practices inferred by analyzing the dataflow using dynamic analysis.  One major limitation of flow-to-consistency analysis is that it requires dynamic analysis of apps which is hard to scale, as pointed out by Xiao et al~\cite{xiao2022lalaine}. 

In this work, we check the consistency of privacy labels using different approaches -- by comparing privacy practices reported in privacy labels with those present in privacy policies. We also compare privacy labels of the same apps across different platforms to gain an understanding of how developers report their apps' privacy practices. A major advantage of our approach is that using automated analysis, it can scale to a large number of apps. Thus, this paper aims to provide a comprehensive analysis of the current state of privacy labels and identify areas for improvement by asking the following research questions:

\begin{itemize}
    \item What practices are developers reporting in privacy labels? How do these practices evolve over time?
    \item How do the privacy practices present in privacy labels compare with the privacy policies? 
    \item Do apps have different practices across platforms?
    % \item Does an app’s privacy policy change since the introduction of the privacy nutrition labels;
\end{itemize}

To answer the above questions, we conduct a large-scale analysis of the privacy labels of apps listed on the Google Play Store and Apple App Store. We also conduct a developer study with android developers to understand their data safety section and highlight the challenges faced by them while working with privacy labels. Our analysis includes a comparison of the privacy practices mentioned in the privacy labels with those present in the privacy policies, as well as a comparison of the privacy practices across apps cross-listed on both platforms. 

We first start by developing a scraper for the Google Play Store and Apple App Store to collect metadata for over 2.5M apps on the Play Store and 1.3M apps on the App Store. We also periodically collected metadata for apps on the play store to track any changes made to an app's description, privacy policies, and data safety section. In addition to collecting each app's metadata, we also scraped privacy policies for apps on both platforms. Next, we automatically analyzed the privacy policies to extract privacy practices by developing a privacy label-centric taxonomy by adapting an existing privacy policy taxonomy. Specifically, we added missing elements and added more annotations for the new taxonomy. 
% We recognize that annotation cost can increase significantly for privacy policies, thus we proposed a bootstrapped approach that leverages advances in Natural Language Processing to minimize the annotation cost. 
We then compare these extracted practices with those present in the privacy label to perform a consistency analysis. Finally, we curate a dataset with apps cross-listed on both platforms and compare the privacy labels to understand how consistent developers are in disclosing their practices via privacy labels.

With this work, we make the following contributions:
% \todo[inline]{Mention the date of data? as in 2.2 as of the last}
\begin{itemize}
    \item We perform large-scale measurements of privacy practices reported in privacy labels across two major platforms - App store (n=\nnumber{1.38M}) and Google Play Store(n=\nnumber{2.4M}). We filter out apps with less than 1000 downloads for Google Play Store. This limits the number of apps on the Google Play Store to \nnumber{1.14M}.  We find that only \nnumber{50.2\%} of the apps provide privacy labels on the Google Play Store, whereas on the App Store, only \nnumber{69.2\%} of the apps contain privacy labels.
    \item We perform a longitudinal analysis for privacy labels on the Google play store and study the evolution of Data Safety Forms before and after the hard deadline imposed by Google. We find that app developers have changed data safety forms frequently. 
    \item We compare the data practices mentioned in the privacy policy with privacy labels for apps in app store and google play store and find that on play store, at least 40\% of the apps have inconsistencies. 
    \item We also identify \nnumber{165K} apps cross listed on both the platforms and compare how the practices are reported. Surprisingly, we find that privacy labels for \nnumber{51.5\%} of the apps are not consistent across the different platforms.
    \item We provide first large scale datasets for privacy labels for Android (n=\nnumber{1.14M}) and iOS (n=\nnumber{1.3M}). Further, we also release the new dataset for the newly formed privacy centric taxonomy. Finally, we release a large policy dataset annotated with the privacy centric taxonomy. The datasets will be available after publication. 
\end{itemize}

\begin{figure}[t]
  \centering
  \includegraphics[width=\columnwidth]{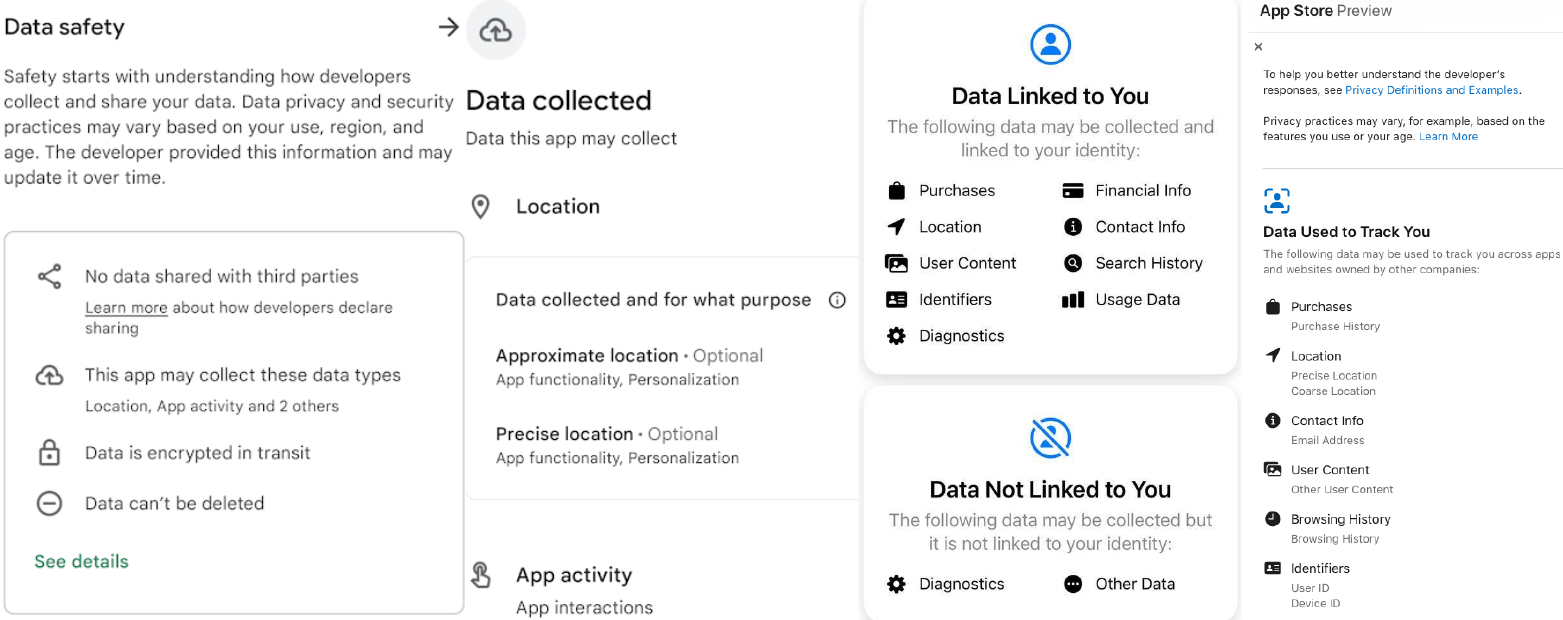}
  \caption{Illustrative Example of nutrition labels}
  \label{fig:pl_example}
\end{figure}

\section{Background and Related Works}
\label{sec:background}

\textbf{Privacy Nutrition Labels.}
Originally introduced by Kelley et al.~\cite{kelley_labels, kelley_study_label}, privacy nutrition labels aim to summarize the privacy practices of websites in a nutrition label format for better visual comprehension. They later designed the ``Privacy Facts'' display to allow the users to consider privacy while installing apps~\cite{kelley2013privacy}. More recently, researchers proposed an Internet of Things (IoT) security and privacy label~\cite{emami2020ask, emami2021privacy} to surface privacy and security information related to IoT devices to the users. Researchers have also studied the design and evaluation of privacy notices and labels~\cite{balebako2015impact, kelley_labels, kelley_privacy_app, kelley_study_label, kelley2013privacy, cranor2012necessary,schaub2015design, mcdonald2009comparative, fox2018communicating, cranor2022mobile}.
\smallskip

In December 2020, Apple adopted the privacy nutrition labels for the app store and mandated that app developers provide their apps' privacy information in the form of the Apple Privacy Label (APL). More recently, Google also required app developers to add a Data Safety Section (DSS) on the Google Play Store.

\begin{figure}[t]
  \centering
  \includegraphics[width=\columnwidth]{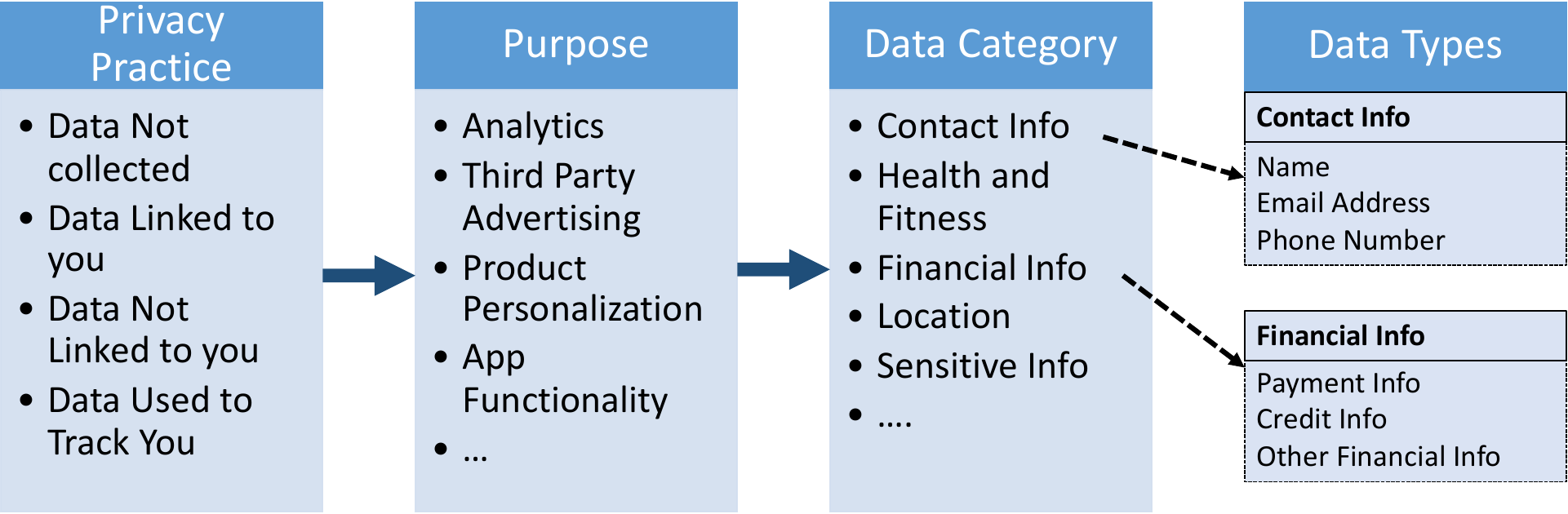}
  \caption{The hierarchy of Apple Privacy Labels}
  \label{fig:apple_label}
\end{figure}

\medskip
\noindent
\textbf{Apple Privacy Label.}
The Apple Privacy Label (APL) is a four-level hierarchy (as shown in \cref{fig:apple_label}). The top level consists of four high-level privacy practices, known as \textit{Privacy Types}. The second level of the label discusses the purpose for data usage, while the third and fourth level describes high-level \textit{Data Categories} and fine-grained \textit{Datatypes}, respectively. In the top level, \textit{No Data Collected} denotes that the app does not collect any data from the users.

Among the other three categories, \textit{Data used to Track you} covers the practices when user data is linked with third-party data for targeted advertising, Ad measurement, or sharing with a data broker. Notably, tracking does not apply when the data is never sent off the device in \textit{a way that can identify the user or device}, or if the data is used for fraud detection. \textit{Data linked to you} covers the personal information and data that is linked to the user's identity as opposed to \textit{Data not linked to you}.

The next level describes the purposes for which data collected in \textit{Data linked to you} and \textit{Data not linked to you} may be used. Apple defines five main purposes: \textit{Third party advertising and marketing}, \textit{Developers' advertising and marketing}, \textit{Analytics}, \textit{Product Personalization}, \textit{App Functionality} and \textit{Other Purposes}. It is important to note that \textit{Data Used to Track you} does not get a purpose level as its purpose is to track the users. In the \textit{Data Categories} level, Apple defines 14 categories of data such as \textit{Contact Info} (consisting of personal information), \textit{Health and Fitness}, \textit{Financial Info} etc. \textit{Data Categories} consists of the final level - \textit{DataTypes} which consists of 32 fine-grained datatypes that the developers can use, such as \textit{App Interactions, Precise Location, Contacts, Phone} etc. An illustrative example of APL is shown in \cref{fig:pl_example}. \smallskip

\noindent
\textbf{Google Data Safety Section}
\label{sec:data_safety}
\begin{figure}[t]
  \centering
  \includegraphics[width=\columnwidth]{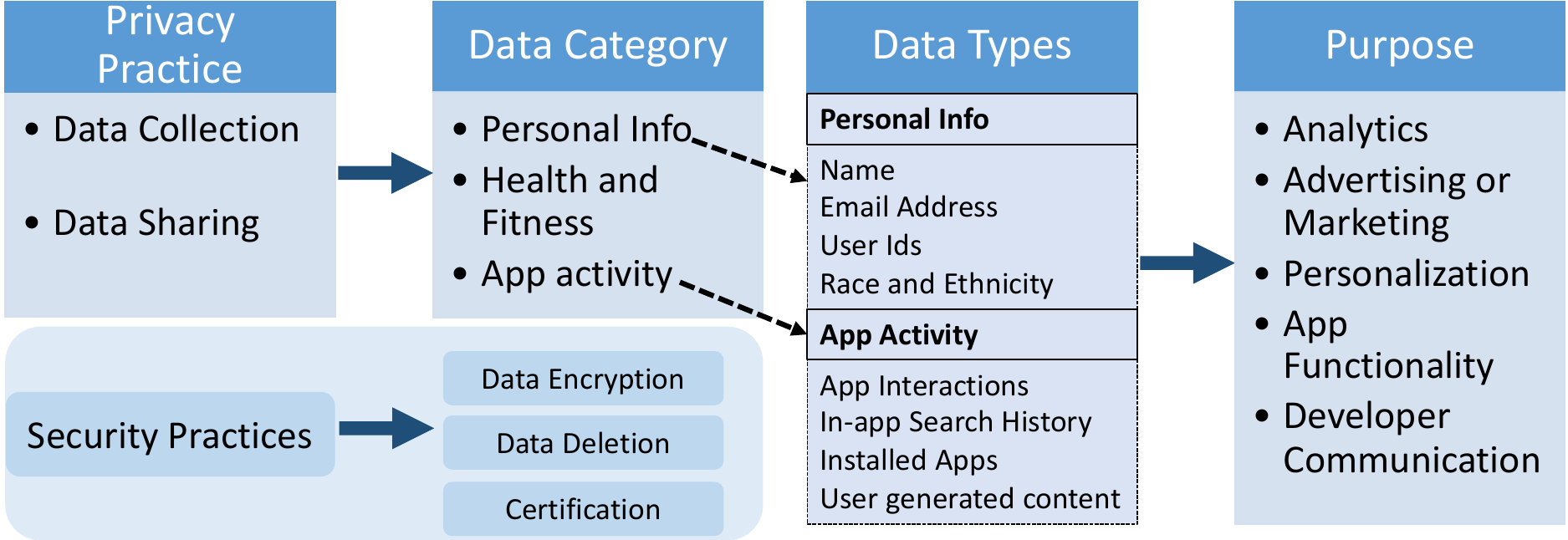}
  \caption{Google Data Safety Section}
  \label{fig:google_dss}
\end{figure}
The Data Safety Section (DSS) also consists of four levels, where the first is high level \textit{Privacy Practices}. The second and third levels consist of \textit{Data Categories} and \textit{Data Types}, and the fourth level consists of \textit{Purpose}.

The first level includes three practices: \textit{Data Collection}, which covers the details about the data that is collected and its intended use; \textit{Data Sharing}, where the developers disclose what data is shared with third parties; and \textit{Security Practices} that covers the data practices related to user choice and data security. \textit{Security Practices} include three tags: \textit{Encrypted in Transit}, \textit{Data Deletion Option}, and \textit{Review against Global Security Standards}.

In the second level, \textit{Data Categories} includes 14 categories such as App Info and Performance and App Activity. Each \textit{Data Category} can also have \textit{Data Types}, which provide fine-grained information about the data used by the app. For example, \textit{App Activity} includes \textit{App Interactions and Installed App}, as shown in \cref{fig:google_dss}. The final level of the Data Safety Section consists of \textit{Purposes} that describe the reasons for collecting or sharing the data.

We note that even though the two privacy labels (APL and DSS) have some overlap at the lowest level, they cover different high-level practices. For instance, APL focuses on surfacing tracking practices and the linkability of the data. DSS focuses on data-centric practices, including collection, sharing, encryption, and deletion. In the rest of the paper, we will use APL and DSS to denote privacy labels for iOS apps and android apps, respectively. Further, we use the term \textit{Privacy Labels} to refer to both APL and DSS collectively.\smallskip

\noindent
\textbf{Usability of Privacy Labels.} Researchers have studied the usability of APLs from both users'~\cite{zhang2022usable} and developers'~\cite{li2022understanding} perspectives. Zhang et al.~\cite{zhang2022usable} studied 24 iPhone users to understand their experiences, understanding, and perceptions of privacy labels on the app store. They uncovered that users find the labels confusing with unfamiliar terms. From the developers' perspective, Li et al.~\cite{li2022understanding} interviewed 12 iOS developers and reported that the sources of errors by developers in privacy labels included both under-reporting and over-reporting data collection. They further concluded that the label design is generally confusing for the developers either due to known factors (lack of resources, improper documentation) or unknown factors (preconceptions, knowledge gaps). More recently, researchers also built and evaluated a tool~\cite{gardner2022helping} that helps iOS developers generate privacy labels by identifying data flows through code analysis. While these works focus on the usability evaluation of APL, our work compares the privacy practices present in privacy policies and labels.\smallskip

\noindent
\textbf{Studies on Privacy Labels.} 
Similar to our work, Xiao et al.~\cite{xiao2022lalaine} characterize non-compliance of apple privacy labels by studying data flow to label consistency of 5K iOS apps. They also provide insights for improving label design. This work is complementary to ours as we measure the consistency of privacy labels with the data practices mentioned in the apps' privacy policies.

The works most similar to ours perform longitudinal measurement of privacy labels to understand the adoption and evolution of apple privacy labels over time~\cite{balash2022longitudinal, li2022understanding, scoccia2022empirical}. In particular, Scoccia et al.~\cite{scoccia2022empirical} conducted an empirical study of 17K apps to characterize how sensitive data is collected and shared for iOS apps. They found that free apps collect more sensitive data for tracking purposes. Li et al.~\cite{li2022understanding} and Balash et al.~\cite{balash2022longitudinal} collected weekly snapshots of apple privacy labels and characterized the privacy practices mentioned in privacy labels for \nnumber{573k} apps. Balash et al.~\cite{balash2022longitudinal} also perform additional correlation analysis with app meta-data like user rating, content rating, and app size.

Our work is different in two ways. First, we provide complimentary analysis by analyzing privacy labels from Apple and Google to provide a comprehensive understanding of practices mentioned in APL and DSS. In doing so, we also verify their findings on how sensitive data is being collected and used. Second, we perform a consistency analysis of privacy labels with privacy policies. We also create a dataset with cross-listed apps on both platforms to understand how developers disclose their practices on different platforms. To the best of our knowledge, ours is the first work performing this analysis. \smallskip

\noindent
\textbf{Automated Privacy Policy Analysis.}
In 2016, Wilson et al.~\cite{wilson2016creation} introduced a privacy policy taxonomy along with an annotated dataset (OPP-115). The taxonomy covers privacy practices mentioned in the privacy policies of the websites. In the past few years, several works have trained classifiers using the taxonomy for automated policy analysis~\cite{harkous2018polisis, srinath2020privacy, wilson2016creation, mousavi2020establishing, wagner2022privacy}. Researchers have also used automated policy analysis to check for consistency within the policy~\cite{andow2019policylint, andow2020actions}, as well as consistency with the code~\cite{zimmeck2016automated, zimmeck2019maps}. Finally, automated analysis has also been used to study the impact of law and regulations on privacy policy~\cite{linden2018privacy, zaeem2020effect}. In this work, we extend the OPP-115 taxonomy to cover practices from nutrition labels and develop new classifiers to extract relevant privacy practices from the policies.

\section{Data Collection Pipeline}
\label{sec:measurement}
\begin{figure}[t]
  \centering
  \includegraphics[width=\columnwidth]{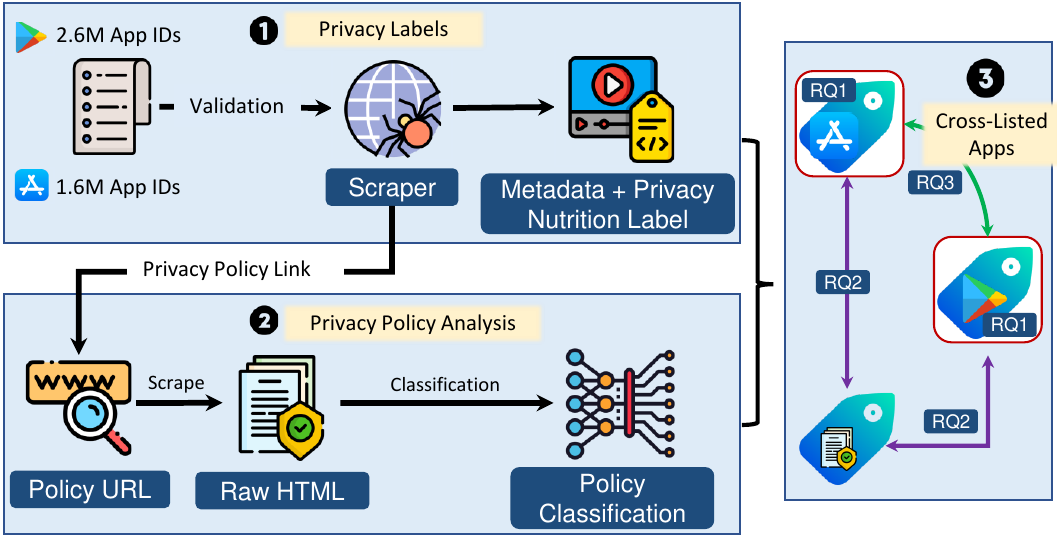}
  \caption{Overview of the data collection pipeline. RQs here refer to the \textit{Research Questions} introduced in \cref{sec:intro}}
  \label{fig:meas_pipeline}
\end{figure}

We show an overview of the data collection pipeline in  \cref{fig:meas_pipeline}. We begin by scraping the metadata and privacy labels for the apps from Google Play and Apple App Store (\cref{sec:privacy_labels}). We then design a classification pipeline to automatically annotate the policies of the mobile apps (\cref{sec:privacy_policy}). Finally, we identify cross-listed apps between Google Play and Apple App Store (\cref{sec:cross_apps}).

% In the following sections, we use the resulting datasets to answer our \textit{Research Questions (RQs)}.
% \todo[inline]{All apps with DSS have PP except 13. there are 994K apps with DSC}

\subsection{Privacy Labels}
\label{sec:privacy_labels}
First, we describe the collection method for our privacy labels (both DSS and APL) datasets.\smallskip

\noindent\textbf{Google Data Safety Section.}
\label{sec:google_dsc}
We collected 10 snapshots of the Data Safety Sections for 2.6M apps present on the play store between June 20, 2022, and Nov 25, 2022. Google required app developers to complete the data safety section by July 20, 2022. By collecting data before and after this date, we are able to capture how the app developers responded to Google's requirement for adding a data safety section to their apps. Note that we captured weekly snapshots from June 20 to Aug 1, which includes the date set by Google. The remaining two snapshots were taken on Sept 9 and Nov 25. 
 
To collect the data safety section, we start with the apk list provided by Androzoo~\cite{androzoo}. This daily updated list consists of up-to-date Android app ids from various sources, including those from the Google Play store. Using the app ids and a 
% Using the app id, we retrieve the app's package name and construct the app URL on Google Play. Then,we use a 
customized version of publicly available google play store scraper library~\texttt{google-play-scraper}~\cite{google-play-scraper}, we capture the metadata of each app, including its data safety sections and the link to the privacy policy.  We used four local machines to perform the scraping. The total time to retrieve data for \nnumber{2.6M} apps, from Google Play, is between 24 to 48 hours. We note that this set also includes apps with very low download counts. To ensure that our statistical analysis is not skewed by these apps, we filter out apps that have fewer than 1000 downloads resulting in a total of \nnumber{1.14M} apps with \nnumber{573k} having privacy labels. We refer to this dataset as \textbf{DSS Dataset}.

Performing the longitudinal analysis, we find that during the period of June 20, 2022, to November 25, 2022, the number of apps with DSS increased from 28.76\% to 47.71\%. The largest change was observed between July 13, 2022, and July 26, 2022, when the percentage of apps went from 35\% to 37.4\% in 13 days. Interestingly, we find that 5\% of apps removed their DSS over the course of our data collection. Specifically, \nnumber{27K} updated their play store page to remove the privacy label. Off these apps, \nnumber{8.6K} apps were deleted, including \nnumber{1.3K} apps which had over 100K downloads. For example, ~\textit{Sport Prediction} with 1M app download had a DSS as of August 1, 2022, but did not have it by Nov 25, 2022.  Furthermore, we find that \nnumber{1.3K} apps updated their DSS to reflect a change in their practices \textit{i.e.} they updated DSS to add or remove privacy practices. We investigate the factors responsible for these changes in \cref{sec:developer}.

\smallskip
%and \nnumber{598k} unique privacy policies

% \kassem{missing stats here}

% The response object from the scraper contains the practices mentioned in the data safety section as well as the meta data such as app name, number of installs, privacy policy URL, app developers' website etc.  Finally, we store the parsed data in a database for further analysis. The privacy policy URL for each app is stored separately which is later used to create the privacy policy dataset (\Cref{sec:privacy_policy}). 

% . The customization\footnote{Note that the library was later updated in August to include the data safety section} was required 

% We use the list to extract the package names for the apps. The package names are unique and can be used to construct the apps' page on the play store. For example, the URL for any app can be constructed by appending the package name to ``\url{https://play.google.com/store/apps/details?id=}''. 
% % Additionally, to filter out the apps that not available on play store, we perform a validation step by checking the existence of the URL constructed using the package name.

 % We note here that Google required app developers to complete Data Safety Section by July 20, 2022. 

\noindent\textbf{Apple Privacy Labels.} 
\label{sec:apple_label_methodology}
In this work, we do not perform a longitudinal analysis for APLs as Balash et al~\cite{balash2022longitudinal} performed a similar study earlier in 2022. Instead, we collect a single snapshot of the Apple Privacy Labels (APLs) on November 13, 2022, for consistency.
To curate the dataset, we begin by parsing the XML site map for the app store.\footnote{We  used the \texttt{ultimate-sitemap-parser} library.} Using the URLs from the sitemap,
we use the Apple Store Catalogue API to extract the metadata for each app, including the privacy nutrition label and a link to the privacy policy. We performed the crawl using 11 instances of google cloud functions to scrape \nnumber{1.6M} apps in 15 hours.

We extracted information for 1.38M apps out of the 1.6M apps available, filtering out those with non-English content. As a result, we obtained 955K (69.2\%) with APLs. In comparison, Balash et al~\cite{balash2022longitudinal} in March 2022 found that 60.5\% of apps had Apple Privacy Labels. The higher percentage in our study suggests that new APLs are still being added to apps. We refer to this dataset as \textbf{APL Dataset}.

% The API response is a JSON string that contains the privacy nutrition label, including the privacy types, data category, data types and purpose. The JSON response also contains the associated meta data of the app such as developer website, app name, privacy policy URL, genre etc. In addition to the API calls, we used the unique URLs to perform headless crawls to extract the privacy policy link shown on the app store page.  

% We leverage cloud functions to perform the API calls to obtain the privacy nutrition labels. The API call can occasionally result in errors if too many requests are sent. To avoid overwhelming the servers, we wait for 2 hours before retrying the failed instances. Using   the scraping process finished in 14-16 hours. 

% Using the privacy policy URLs extracted during privacy labels scraping, we retrieve the raw HTML of the policies and extract the cleaned text. We then pass the extracted text to a classifier that determines whether the webpage is a privacy policy. Although  both platforms require app developers to add a privacy policy link to their app, we found that the privacy policy URL is either unavailable or is invalid in many cases (more details later). Then, we develop a new taxonomy for privacy policies that accommodates the privacy labels. We train a suite of classifiers that annotate a privacy policy using labels from the developed taxonomy. 
% \subsubsection{Privacy Policy Dataset} 

\subsection{Privacy Policy Analysis}
\label{sec:privacy_policy}
We build a privacy policy analysis pipeline to automatically annotate the privacy policies of the apps (the second component in \cref{fig:meas_pipeline}). This annotation allows us to analyze the consistency between the privacy labels and privacy policies of each app at a scale (\cref{sec:policy_inconsistency}).\smallskip

\noindent
\textbf{Text Extraction and Cleaning.}
Starting with the privacy policy URL, we crawl the corresponding webpage, clean the HTML by removing the headers and footers, and extract the text using the \texttt{BeautifulSoup} library~\cite{richardson2007beautiful}. We use the \texttt{PyPDF2}~\cite{pypdf2} library to extract text from privacy policies that are PDF documents. Following prior works~\cite{harkous2018polisis}, we apply three exclusion criteria. First, we filter out the instances where the text length is less than 100 words. Second, we filter out the instances where the policy is stored in non-standard formats, such as images. Finally, as our analysis pipeline relies on the English language, we filter out non-English policies (330K for Play Store, 230K for App Store) using the language detection library \texttt{polyglot}~\cite{polyglot}.\smallskip

\textbf{Policy Classification.}
To extract the practices from privacy policies, we follow the pipelines from existing works~\cite{harkous2018polisis, linden2018privacy} that use OPP-115 taxonomy~\cite{wilson2016creation}. However, in our case, not all practices mentioned in privacy labels can be extracted as the underlying taxonomy either lacks these classes or the datasets do not have sufficient samples. For example, a data category used in APL, \textit{Sensitive Info} is missing from the taxonomy. To overcome this limitation, we extend the existing taxonomy by adding missing classes. Further, two of the authors perform annotation for the added classes. The annotators had an overlapping set of 200 segments to measure the inter-annotator agreement. We find that the annotators showed high agreement with a Cohen's Kappa value of ($\kappa = \red{.85})$. We provide more details about the taxonomy and the annotation in \cref{appendix:taxonomy}.

Following prior works~\cite{wagner2022privacy, srinath2020privacy}, we use DistilBERT~\cite{sanh2019distilbert} to train the classification models. We then use these models to extract privacy practices from policies. This approach allows us to accurately extract the practices from privacy policies, even for those practices that were not covered by the existing taxonomy. \cref{tab:policy_results} shows the performance of our classifier on a held-out test set. We note that our classifiers outperform previous classifiers, primarily because we added new annotations. 
%A full comparison for classes can be found in \cref{appendix:taxonomy}.

\begin{table}[t]
\centering
\begin{tabular}{lccc}
\toprule
\textbf{Category} & \textbf{CNN ~\cite{o2015introduction}} & \textbf{BERT ~\cite{devlin-etal-2019-bert}} & \textbf{Ours}\\
\midrule
\rowcolor{aliceblue}First-party-collection-share  & 82 & 91 & \textbf{98} \\
Third-party-sharing-collection & 81 & 90 & \textbf{96} \\
\rowcolor{aliceblue}Identifiability & 77 & 91 & \textbf{97}\\
Does-does-not & 86 & 93 & \textbf{96} \\
\rowcolor{aliceblue} Encryption-in-transit  & N/A & N/A & \textbf{99} \\
Data Deletion Option  & N/A & N/A & \textbf{91} \\
\bottomrule
\end{tabular}
\caption{Selected Classifiers' performance on the test set. For the performance of all classifiers, please see \cref{tab:privacy_label_classifier_stats} in  \cref{appendix:taxonomy}.}
% \todo[inline]{Paul: Update CNN and BERT stats}
\label{tab:policy_results}
% \vspace{-8mm}%
\end{table}

\subsection{Identifying Cross-Listed Apps}
\label{sec:cross_apps}
We next describe the process to identify cross-listed apps across the two platforms ((3) in \cref{fig:meas_pipeline}). We use the resulting dataset to compare the privacy labels of apps across the two platforms (\cref{sec:consistency_cross}). Identifying two versions of the same app across platforms is challenging due to the lack of unique identifiers~\cite{hooda2022skillfence}.

To uniquely map apps across platforms, we develop a heuristic based on combinations of pseudo-identifiers, such as the app name, developer name, privacy policy, and developer website. We start with the apps which have the same name across both platforms (n=\nnumber{220K}). Next, if the privacy policy of the apps matches, then we treat them as a unique match (n=\nnumber{85K}). In some cases, like the \textit{NTLC Catalog} app on the Play Store and App Store, the app developers can include platform-specific identifiers in the URLs for privacy policies. To capture these instances, we match the first level domain of the privacy policy URLs and identify them as unique matches (n=\nnumber{54K}). Finally, while providing privacy policy links is highly encouraged in both platforms, some apps do not contain the link to the privacy policy. To further increase the coverage, we also match the first-level domain of the developer website, which is present on both platforms. Using these criteria, we are further able to get \nnumber{25K} matches. This way, we obtain a total of \nnumber{165K} apps that have instances in both Apple play store and Google play store.\smallskip

\noindent
\textbf{Manual Verification:} To assess whether our heuristic results in false positive matches, two of the authors manually verified 150 app pairs identified using each of the three heuristics and found that no app from Google Play Store was matched to an incorrect app from App Store. It is worth noting that for our analysis, having an accurately mapped set is more important than capturing all instances of cross-listed apps. \smallskip

\noindent
\textbf{Cross-listed Apps Dataset}
Using the method described above, we find a total of \nnumber{165K} cross-listed apps. Among these apps, we find that \nnumber{5\%} have privacy nutrition labels only on the Google Play Store, \nnumber{20.2}\% have the label only on the Apple App store, \nnumber{60.8}\% have labels on both the platforms and \nnumber{13.9}\% do not have a privacy nutrition label on either platform. The higher rate of privacy labels for the App store can be understood as Apple enforced nutrition labels on their platform earlier than Google, giving more time for developers to add the details in the APL.

\subsection{Ethical Considerations}
We collected data only from publicly available web pages and APIs. While our data collection scripts might load Google and Apple's servers, we were careful to not abuse these resources. In particular, we added back-off strategies in case of errors and waited for sufficient time before retrying for the failed cases. Furthermore, for privacy policy extraction, we were respectful of robots.txt and only extracted HTML when the website allowed us to.

\subsection{Observations}
Our measurement pipeline results in two initial observations. First, app developers have been slow to add privacy labels to their apps, even after the hard deadlines have passed. Privacy labels are present only for 69\% of the apps on the Apple app store and \nnumber{50.2}\% of the apps on the Google play store (as of November 2022). Second, our measurement pipeline produced large-scale datasets for Apple Privacy Label (n=955K), Google Data Safety Section (n=\nnumber{573K}), and privacy policies (n=\nnumber{598K}) corresponding to the apps. In addition, we generated a new \textit{Privacy Label Taxonomy} by adding missing elements to OPP-115 taxonomy. We also supplement the existing privacy policy datasets by adding annotations for the new categories.

\section{Data Practices in Privacy Labels}
% In this section, we study how the app developers report their apps' privacy practices in the privacy labels. Specifically, we use the APL and DSS datasets curated in \Cref{sec:measurement} to answer the question: \textit{How do applications collect and use data?}. Further, we analyze the privacy labels for apps along three dimension: age rating, price and popularity. Finally, we identify sensitive data flows and perform a case study to show how apps can misuse the data.
\subsection{Google Data Safety Section}
\label{google-data-safety-sec4}
In this section, we analyze the DSS dataset (\cref{sec:google_dsc}) comprising \nnumber{573K} apps. We first discuss the practices present in DSS and then examine how these practices vary with an age rating, price, and popularity.
% In this section, we begin by discussing the high level practices. We then discuss the practices at \textit{Data Category} and \textit{Purpose} level. In our  we find that \nnumber{50}\% of the apps on the play store have Data Safety Sections.\smallskip

\begin{figure}
  \centering
  \includegraphics[width=\columnwidth]{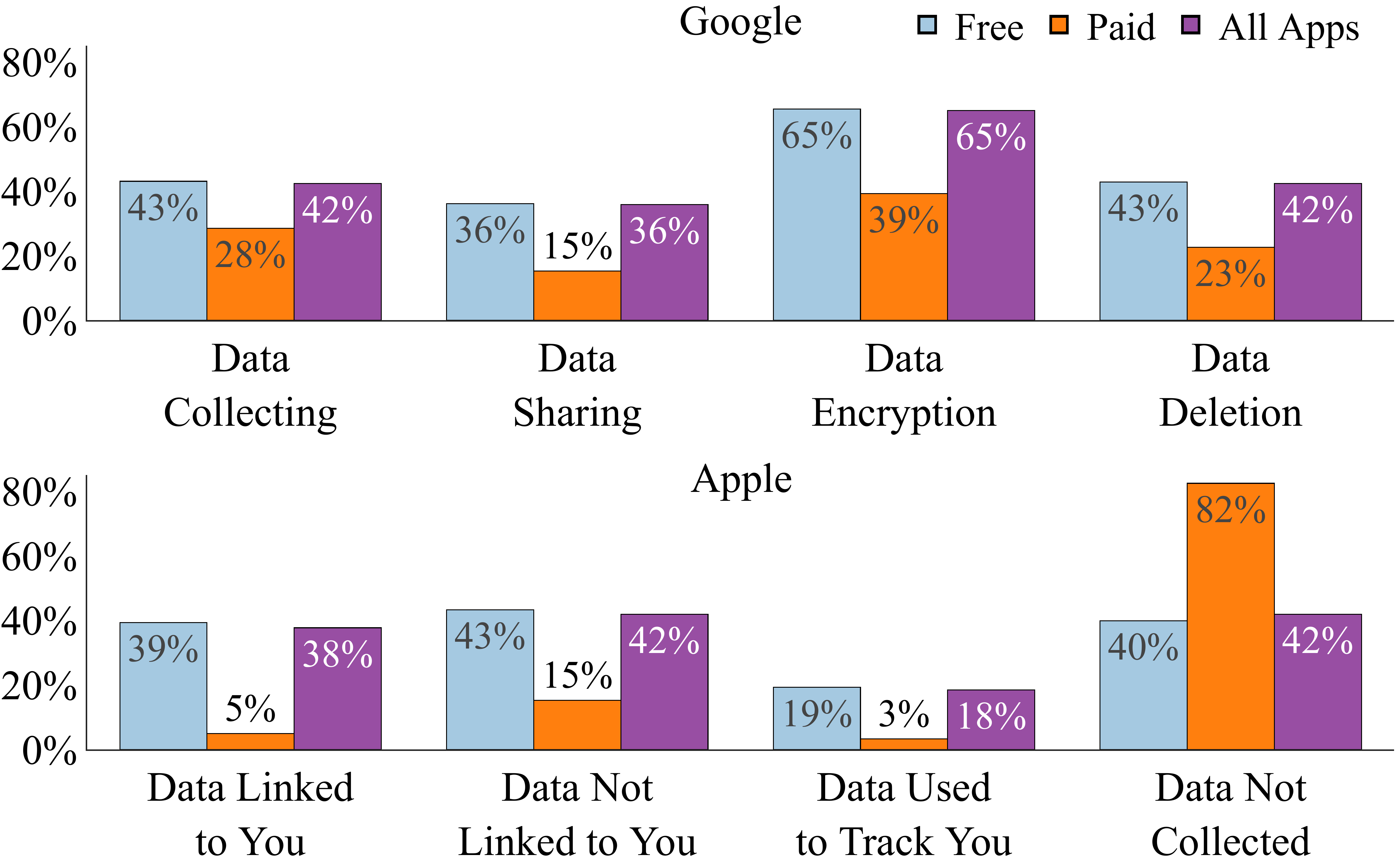}
  \caption{Distribution of privacy types in Google Data Safety Sections and Apple Privacy Labels. The normalization is done by the total number of apps with privacy labels.}
  \label{fig:free_paid}
\end{figure}

\noindent
\textbf{Data Collection and Sharing:} 
Among the apps having DSS, we saw \nnumber{42.3}\% collecting at least one type of data, and \nnumber{35.8}\% sharing at least one data type (purple bars in the top plot for \cref{fig:free_paid}). This suggests that the majority of the apps on the play store report do not collect or share data. This is in contrast with the findings from prior work~\cite{wang2015wukong} that found that the majority of the apps use at least one third-party application, which has been shown to collect sensitive information~\cite{book2013longitudinal, lin2013understanding}. One possible explanation for this is that developers find it hard to understand the collection and sharing practices of third-party libraries. This is also supported by prior research~\cite{balebako2014privacy, li2022understanding}. As such, when inquired about change in DSS, one developer also alluded to lack of transparency by third parties:\newline
\textit{``We don't collect or share any user data. But we use Meta (former Facebook) audience network for monetizing non-paying users with ads. Unfortunately, the details provided by Meta are very vague..''}

We also note that among the apps not collecting any data around \nnumber{23}\% are sharing data. This is because \textit{Data Collection} is defined as the instance when the developers retrieve the data from the device using the app~\cite{googledocumentation}, whereas \textit{Data Sharing} is defined as when the data is transferred from the device to a third party. This way, the developers can share data without collecting it if the application uses third-party libraries which directly send data to their servers.\smallskip

\noindent\textbf{Security Practices:}
We find that \nnumber{23}\% of the apps do not provide any details of their security practices. \nnumber{65}\% of the apps encrypt data that they collect or share while it's in transit, and \nnumber{42}\% allow the users to request that their data be deleted or automatically anonymize data within 90 days. Notably, we find that \nnumber{17.4\%} of the apps state that they do not collect or share data, but encrypt the data in transit. We explore this behavior further in \cref{sec:developer}. As apps need network permissions to transmit data, we cross-verified encryption practices with apps' network permission requests and find that \nnumber{10.5\%} apps do request network permission but do not encrypt data, potentially exposing user data in plain text. Additionally, \nnumber{2.2\%} of apps do not request network permissions, yet state that they encrypt data in transit, suggesting that some developers might be over-reporting their practices, consistent with prior research~\cite{li2022understanding}.\smallskip

\begin{figure}
  \centering
  \includegraphics[width=\columnwidth]{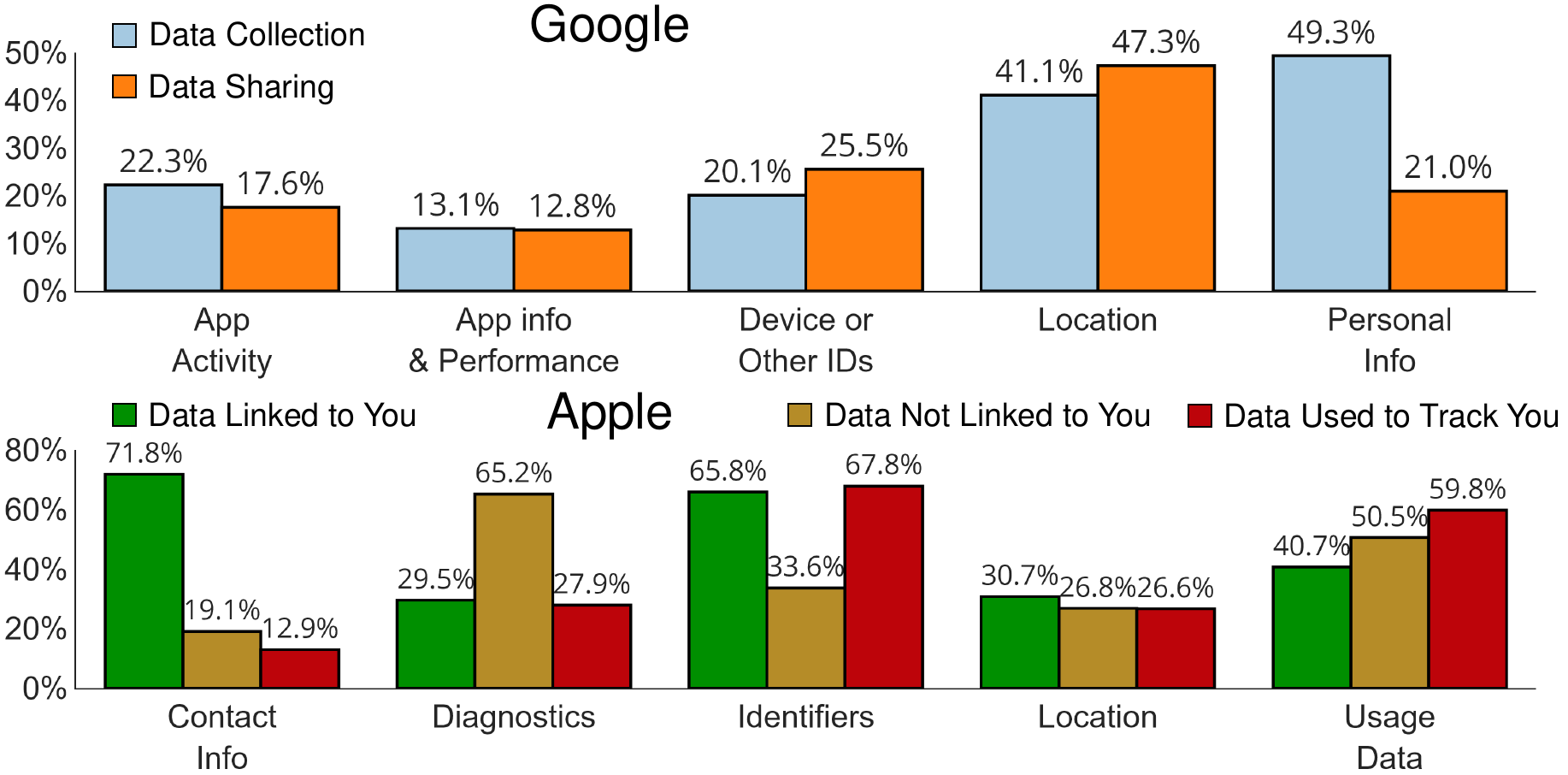}
  \caption{Distribution of Top-5 data categories for high-level practices for apps in Play Store (top) and App store (bottom). The normalization is done by the total number of apps with privacy labels. For plots with data categories, see \cref{fig:inconsistency_app_no_thresh} in the Appendix.}
  \label{fig:google_collected_sharing}
\end{figure}

%and among the apps not sharing any data, 30\% are collecting data. 

\noindent\textbf{Category and Purpose Level Practices:}
In \cref{fig:google_collected_sharing}, we present the top-5 data categories for \textit{Data Collection} and \textit{Data Sharing} by apps in play store. A full plot including all data categories can be found in \cref{fig:inconsistency_app_no_thresh} (Appendix).
Our findings indicate that the data categories \textit{Personal Information} and \textit{App activity} are among the most frequently collected, and are primarily used for \textit{App functionality} and \textit{Analytics}. However, \textit{Location} and \textit{Device Ids} are more commonly shared for the purpose of \textit{Advertising or Marketing}. We emphasize that this flow poses serious privacy risks and allows for tracking by third parties.
We also observe that sensitive data types such as \textit{Audio}, \textit{Files and Docs}, and \textit{Health and Fitness} are collected less frequently, with the most common purpose being \textit{App functionality}. Furthermore, we note that out of the 7 possible purposes for collecting data there are over \nnumber{4K} apps that list 6 or more purposes for the data they collect, which may indicate that app developers list all purposes out of convenience. For example, \textit{Workplace from Meta} with over 15M+ downloads, lists the same 6 purposes for all the data they collect like access to \textit{Installed Apps}, \textit{SMS or MMS}, \textit{Music Files}. This is consistent with the findings of Li et al.~\cite{li2022understanding}, who suggest that developers may over-report in cases of ambiguity.\smallskip

\noindent\textbf{Variation of Practices with Popularity:}  We first investigate the relationship between privacy practices and app popularity. We classify apps into three categories based on their number of downloads: extremely popular (greater than 1M download, n=56K), semi-popular (more than 10K downloads, n=524K), and low-popular (less than 10K downloads, n=621K).
Our findings reveal that 1) the fraction of apps displaying Data Safety Sections (DSS) increases with the popularity of the apps (42\% for low-popular, 51\% for semi-popular and 76\% for extremely popular) and 2) the fraction of apps collecting and sharing data is less for popular apps (41\% for low-popular, 46\% for semi-popular and 12\% for extremely popular). These results suggest that developers from popular apps tend to report more privacy-friendly practices.\smallskip

\begin{figure}
  \centering
  \includegraphics[width=\columnwidth]{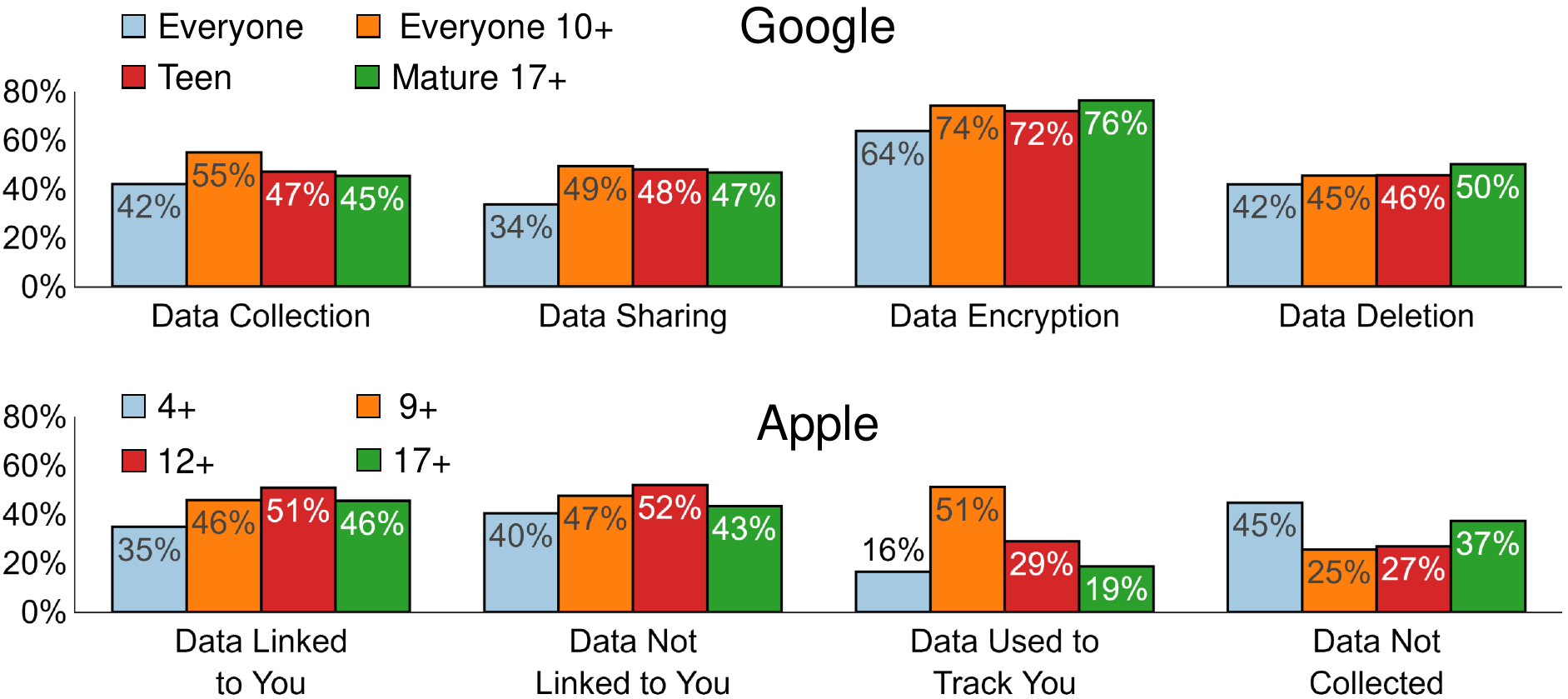}
  \caption{Distribution of privacy types based on age rating for DSS and APLs. The normalization is done by the total number of apps with privacy labels.}
  \label{fig:age_rating}
\end{figure}

\noindent\textbf{Variation of Practices with Age Rating:}
Next, we examine how the practices of apps differ based on their age rating as determined by the Google Play Store. The Play Store assigns five different age ratings: Everyone, Teen, Mature 17+, and Everyone 10+\footnote{Google also has Adults 18+ rating, but we found less than 200 apps in this category and decided to filter it out for this analysis}. We acknowledge the importance of this distinction, as apps that are accessible to children and teens (falling in the Everyone and Teen categories) are expected to have higher transparency and collect less data. However, our analysis of the dataset reveals that 59\% of apps with the \textit{Mature 17+} rating have a Data Safety Section (DSS), while the fraction of apps with a DSS in the other age ratings ranges from 47\% (Everyone) to 55\% (Everyone 10+). The data practices for different age ratings are shown in \cref{fig:age_rating}.  We find that the fraction of apps having \textit{Data Collection} and \textit{Data Sharing} is lowest for apps rated for \textit{Everyone}, whereas apps targeting \textit{Mature 17+} have the highest encryption rate.\smallskip

\noindent\textbf{Variation of Practices with Price:}
Finally, we study the difference in practices based on whether the app is available for free, free with in-app purchases, or paid. We find that 68\% of the paid apps have DSS whereas, for free apps, only 46\% have DSS. \cref{fig:free_paid} shows the distribution of high-level practices with free and paid apps. We note that for paid apps, a fraction of apps collecting and sharing data is lower. Furthermore, apps with \textit{Data Encryption} and \textit{Data Deletion} are lower because the apps are collecting and sharing fewer data. This suggests that paid apps tend to have better data practices. 

% \rishabh{@Asmit: Add the detailed analysis for encryption here}.

%=======================================================================================================================

\subsection{Apple Privacy Labels}
Next, we examine the Apple Privacy Label (APL) dataset (\cref{sec:apple_label_methodology}) consisting of privacy labels from \nnumber{955K} apps. We first discuss the practices present in APL and then dive into variations of practices with an age rating and price. Finally, we conclude by comparing the low-level practices mentioned in APL and DSS \smallskip

\noindent
\textbf{High-Level Practices:} In our dataset, 42\% of apps collected data from users that were not linked back to the user (Data Not Linked to You), whereas 37\% of apps did collect data that is linked to the user (\cref{fig:free_paid}. Note that apps could collect multiple types of data some of which may be linked to the users while others may not. Furthermore, around 18\% of the apps reported collecting data that was used to track the users. Note that this reflects the status of the APLs after the \textit{Apple Tracking Transparency} policy was implemented, which requires developers to obtain consent from users before tracking. We also find that 42\% of apps report that they do not collect any data from users. Recent works~\cite{kollnig2021iphones, kollnig2022goodbye} analyzing iOS apps have found that at least 80\% of the apps still use tracking libraries in the apps. Further, these libraries have been shown to collect user data~\cite{book2013longitudinal, lin2013understanding}. Similar to the case of android developers, this discrepancy can be explained by the lack of transparency of privacy practices by the third-party libraries, resulting in confusion for the developers. \smallskip
% \rishabh{Should we go one level deeper and talk about data types as well? }

For \textit{Data Used to Track You}, we find that \textit{Usage Data} and \textit{Identifiers} are most commonly used. We note here that Apple defines \textit{Tracking} as when data collected is linked with third-party data for targeted advertising, as well as when the data is shared with a data broker. Additionally, we observe that 25\% of the apps collecting \textit{Location} information also use it for tracking. This poses severe privacy risks to the users as entities can track the physical location of the users which can reveal sensitive details about users' habits and routines. \smallskip

\noindent\textbf{Data Category and Purpose Level Practices:} In \cref{fig:google_collected_sharing} (bottom), we show the top-6 data categories mentioned in the high-level practices in the APL dataset. We find that for \textit{Data Linked to You}, \textit{Contact Information} and \textit{Identifiers} are collected most frequently, whereas for \textit{Data Not Linked to You}, \textit{Diagnostics} and \textit{Usage Data} are collected most frequently. Apple defines \textit{Contact Information} as name, email, phone number, and physical address, whereas \textit{Usage Data} refers to product interactions and advertising data such as information about the ads that the user has viewed. Analyzing purposes for these data categories, we find that nearly 60\% of the apps use these data categories for \textit{App functionality} and \textit{Analytics}. It is also worthwhile to note that \textit{Contact Information} is used for \textit{Advertisements} in only 8\% of the apps that collect this information, indicating that apps generally do not use personal information for advertisements. We also note that \textit{Identifiers}, commonly used for tracking users for targeted advertising is used for \textit{Advertisement or Marketing} in more than 20\% of the apps that collect \textit{Identifiers}. Interestingly, \textit{Location}, under \textit{Data Linked to You} is also used for \textit{Advertisement or Marketing} by 20\% of the apps that collect \textit{Location}.\smallskip

\noindent
\textbf{Variation of Practices with Age Rating} Next, we investigate the correlation between the privacy practices described in the Android Permission List (APL) and the age rating and price of apps. The App Store assigns four different age ratings: 4+, 9+, 12+, and 17+ (which roughly align with the rating system used by the Google Play Store). Our analysis reveals that the fraction of apps with an age rating of 17+ is highest at 76\%. However, we note that the high-level data practices, shown in Figure 1, are consistently more privacy-friendly for apps with lower age ratings. For instance, only 13\% of the apps with an age rating of 4+ track users. Similarly, data collection for these apps is also consistently lower than that of other categories.\smallskip

\noindent
\textbf{Variation of Practices with Price:} Finally, we categorize the dataset into free and paid apps and examine the differences in privacy labels. Recall that for the play store, we observed that paid apps contained more DSS than free apps. For APL, we find the reverse trend with 70\% of the free apps having APL as compared to 52\% paid apps. On the other hand, the high-level practices are decidedly better for the paid apps, as shown in \cref{fig:free_paid} (bottom chart). For instance, 82\% of the paid apps reported not collecting any data, while only 3\% of paid apps mentioned using data to track the user. This indicates that the paid apps on iOS platforms are more friendly than the free apps.\smallskip

\noindent
\textbf{Comparison Between DSS and APL:} As discussed in \cref{sec:background}, DSS and APL provide different information to the users, and cannot be directly compared based on high-level practices. However, since the underlying data collected is the same, we can compare the practices shown in \cref{fig:google_collected_sharing}. We observe that the fraction of apps requesting similar datatypes is much smaller for apps on the play store than that of the app store (with a notable exception of Location). This can be attributed to the fact that developers have had a longer to work with the APL framework, while the DSS framework is still relatively new. In our communication with app developers, one app developer mentioned that they try different answers on the data safety form. We also received communication indicating that some developers updated their DSS based on the questions that we had asked. This indicates that the developers are unclear on the process involved in the data safety forms, which might result in some inaccuracies in the DSSs. This is also supported by the study conducted by Li. et al ~\cite{lin2013understanding} where they find that app developers find it difficult and challenging to fill out the privacy labels, especially, because the frameworks for Apple and Google are starkly different and can create confusion.

\subsection{Developer Study}
\label{sec:developer}
In \cref{sec:privacy_labels} and \cref{google-data-safety-sec4}, we identified three trends in Data Safety Sections: (A) apps stating that they encrypt data without collecting or sharing data, (B) apps changing their practice from not collecting/sharing data to collecting/sharing data, and (C) apps changing their practices from collecting/sharing data to not collecting/sharing data. To gain a deeper understanding of these trends, we reached out to developers via email and asked them one general question about their Data Safety Sections and one specific question about the type of trend we observed in their app. We contacted 30K developers from the Play Store. It is worth noting that, since Apple does not provide email addresses for developers, we only conducted this study with Android developers.

In our initial email, we clearly identified ourselves as researchers and stated that we were studying their application and wanted information regarding their data safety section (\cref{app:dev-study}). Additionally, we do not collect any personally identifiable information from the developers and only use their publicly available contact information from the Play Store to contact them. As such, the study has been approved by the IRB at our institute.\smallskip

% One of the common themes in all the findings so far has been inconsistency in the privacy labels. With this, a natural question arises: \textit{What are the factors that contribute to this inconsistency?} To understand this further, we reached out to 50K android app developers via email and requested more information surrounding the inconsistencies within the Data Safety Cards. In our initial email, we clearly identified ourselves as researchers and then stated that we were studying their application and wanted information regarding some inconsistency in their data safety section. 
\noindent
\textbf{Findings:} Based on our initial emails, we received 2500 responses. After filtering out the automated replies using keyword filtering, we were left with 889 responses where the app developers describe the challenges they face while working with the privacy labels, as well as provided information about their data safety section. We further manually examined each response and curated a set of 307 replies. This manual filtering removed replies that included non-relevant replies. 
% In this section, we analyze these responses from developers by manually coding the responses. 
Next, one of the authors manually coded the responses to identify the factors for the trends, as well as general challenges described by the developers. Another author independently verified the findings by coding a subset of 50 responses independently. Specifically, we first present the major contributing factors for the different types of trends mentioned above. We then discuss the top challenges that developers face while working with the Data Safety Section.\smallskip
% Our approach has the natural advantage of extracting information under natural working environment, as opposed to survey based studies or interview based studies where the participants have the notion \rishabh{Complete this section once the analysis is complete.}

\noindent
\textbf{Type A: Apps Stating that they encrypt data without collecting or sharing data}: For this trend, we obtained responses from 165 developers. Of these, 56\% mentioned data is collected by third-party services like ads or Google Firebase but were not sure if it should be added to DSS while another 36\% were not sure what data was collected, 3\% of the developers were confused regarding encryption and added the option thinking of SSL encryption for communications between the server and the app, without collecting/sharing data. For example, one of the developers said the following: \textit{``I use Google's own libraries for this. In the Google Play Console, Policies section, I had to guess that Google is sending data and I rely on Google to encrypt that data. Because Google says that the developer is responsible for the libraries they use. That's why you find the contradictory result.''}\smallskip

\noindent
\textbf{Type B: Apps changing their practice from not collecting/ sharing to collecting/sharing:} For this trend, we received responses from 130 developers, 12\% of whom did not understand the process and selected any option that was accepted whereas 74\% changed DSS after realizing that third party libraries are collecting data. 12\% of them had an app update while 2\% changed DSS to ensure that they were up to date with the regulations like GDPR. For example, one developer said, \textit{``... Admob SDK I am integrating with the app might collect information [...] And According to Google policy, if I am using the latest version of their Admob SDK, I have to specify that the app is collecting or sharing data ...''}\smallskip

\noindent
\textbf{Type C: Apps changing their practice from collecting/sharing to not collecting/sharing:} 
We only obtained 12 responses for this trend, 58\% of which stated that their app was updated, but the DSS reflects an older version, 25\% mentioned that data was collected by ad libraries that have since been removed, and 9\% mentioned regulations as a factor for the change in DSS. For example, one developer said, \textit{``...we have changed the data safety section of our application because we [...] removed any data collecting libraries such as Firebase [...] Admob for monetization...''} \smallskip

\noindent
\textbf{Challenges for Developers: } We find that the developers are generally confused about how to fill the Data Safety Section. The source of confusion varies from \textit{Not understanding the Process} to \textit{Not understanding if data collected by the third party should be reflected in DSS}. For example, one developer stated 
\textit{``What [...] keeps changing every few months is Google's privacy policies.  They are difficult to understand and they shift like sand... I don't really understand half of them and so we just keep submitting answers in hopes it's what they are looking for...''}
indicating that the process is very unclear, while another mentioned \textit{``the reason for the change was because google play forced me to put that information''}. These confusions are problematic as they may result in inaccurate privacy labels. They can also under-represent the privacy practices in the privacy labels which can give a false sense of security to the users, increasing their privacy risks. 

We note that in an earlier qualitative study, Li et al.~\cite{li2022understanding} found that ``Developers felt unconcerned about privacy and that it was not their responsibility''. In our study, we found that developers cared about user privacy, but did not have enough means (either lack of resources or lack of transparency with third-party libraries' privacy practices) to create accurate labels, causing some frustration on their part. For example, one developer said \textit{``... we use Meta (former Facebook) audience network for monetizing non-paying users with ads. Unfortunately, the details provided by Meta are very vague, but definitely are considered as collecting and sharing data. If possible we would love to switch to an ad provider that offers proper non-personalized ads with zero/minimal data collection, but it seems impossible to find such a provider.''}. 

\subsection{Takeaways}
The analysis presented here results in three main takeaways: 1) Privacy practices reported in the privacy nutrition labels differ from the privacy practices derived using app analysis by prior works~\cite{wang2015wukong}. Specifically, prior works have shown that third-party libraries are used in the majority of the apps and that these libraries collect sensitive information from the users. This is inconsistent with what we find in the privacy labels. This inconsistency can be explained by the fact that privacy practices of third-party libraries are often vague and create confusion among the developers (consistent with findings from literature~\cite{li2022understanding}). 2) We also show that paid apps, and apps that are open to all age groups, including children, are more privacy-friendly. As shown in \cref{fig:free_paid} and \cref{fig:age_rating}, these apps are less likely to engage in tracking, data collection, and data sharing. 3) \cref{fig:google_collected_sharing} also shows that location data is often used for advertising, marketing, and tracking. This poses severe privacy risks, as location data can reveal sensitive information about an individual's habits and routines. Our research suggests that further attention should be paid to the use of location data in mobile apps, and the potential risks it poses to user privacy.

\section{Practices Present in Privacy Policies}
\label{sec:policy_inconsistency}
The next research question that we answer is: \textit{How do the privacy practices mentioned in the privacy labels of the apps compare with the privacy practices described in their privacy policies?} We perform this comparison by training machine learning classifiers to automatically extract privacy practices mentioned in the privacy labels, as described in \Cref{sec:privacy_policy}. For Google's DSS we have \nnumber{346K} apps with valid policies, whereas for Apple, we have \nnumber{343K} apps. As described in \cref{sec:privacy_policy}, we filter out the policies which are not in English. 
% \rishabh{@Rishabh: Make sure that this is well described in earlier section and then adjust text accordingly here.} 
Note that to obtain presence of a particular practice in privacy policy, we require that there exists at least one segment which classifies the segment for that practice. For example, for a policy to have \textit{Data Encryption} practice, we require presence of at least one segment where our classifier tags it as positive for \textit{Data Encryption}. A complete mapping from classifiers to practices in APL/DSS is described in \cref{tab:policy_to_label}.

% It is important to note here that we do not rely on high level classifiers such as \textit{First Party Collection} or \textit{Third Party Data Sharing} for extracting high level privacy practices such as \textit{Data Collection} and \textit{Data Sharing}. This is done to avoid generic collection or sharing statements contaminating the 

There are two types of inconsistencies that can arise: 1) \textit{In Label}, where a given privacy practice is mentioned in the privacy label but is absent from the privacy policy, and 2) \textit{In Policy}, where a practice is found in privacy policy but is missing from privacy label. As privacy policies can potentially cover multiple applications, websites and products, \textit{In Policy} inconsistency does not necessarily mean that policy is inconsistent with the privacy label. For example, the Google app \textit{Clock} reports that it does not collect or share any \textit{Location} information. However, since Google has one policy to cover all the products, the policy states that they can collect \textit{Location} (applicable in Google Maps). In such cases, it is inaccurate to say that privacy label are inconsistent with the privacy policy without further analyses. 
However, if an app mentions collection of data and the policy does not mention it, then we can conclusively say that the privacy policy and the app are inconsistent. Thus, in this work, we will focus primarily on \textit{In Label} inconsistencies, except when there is a negative practice is involved (\textit{Data Not Collected} or \textit{Data Not Linked to You}). This is because if the policy says that data is not collected, then no app corresponding to that policy should collect any data, and in this case we focus on \textit{In Policy} consistency.

% \red{The discussion above also highlights a major issue with using privacy policy as the source to understand privacy practices of the app. Specifically, we highlight the notion that privacy policies usually cover practices for websites (and potentially other products) and hence, may not accurately inform users about privacy practices of the application. In fact, they can mis-inform the user about the privacy practices in certain cases. \rishabh{Maybe move this to discussion and add that while privacy labels address this issue, without consistency checks, they do not contribute much.}}

% \subsection{Comparison with Privacy Labels}
% In this section, we look at how the privacy policies of apps compared with their respective privacy labels as stated in DSS and APL. For Google's DSS we had \nnumber{346K} apps and for Apple, we analyzed \nnumber{458K} apps. \rishabh{The total number of apps with DSS/APL is way higher, right? So these are the apps which had both DSS/APL and privacy labels? If so - are we saying that 350K apps on google had DSS but not a valid privacy policy? This would needs more explanation}\red{These numbers represent the number of apps that have both: privacy labels and a classified privacy policy. Here a classified privacy policy is defined if (1) it exists, (2) is in English.}\smallskip
\noindent
\subsection{Google Data Safety Sections}
\begin{figure}
  \centering
  \includegraphics[scale=0.25]{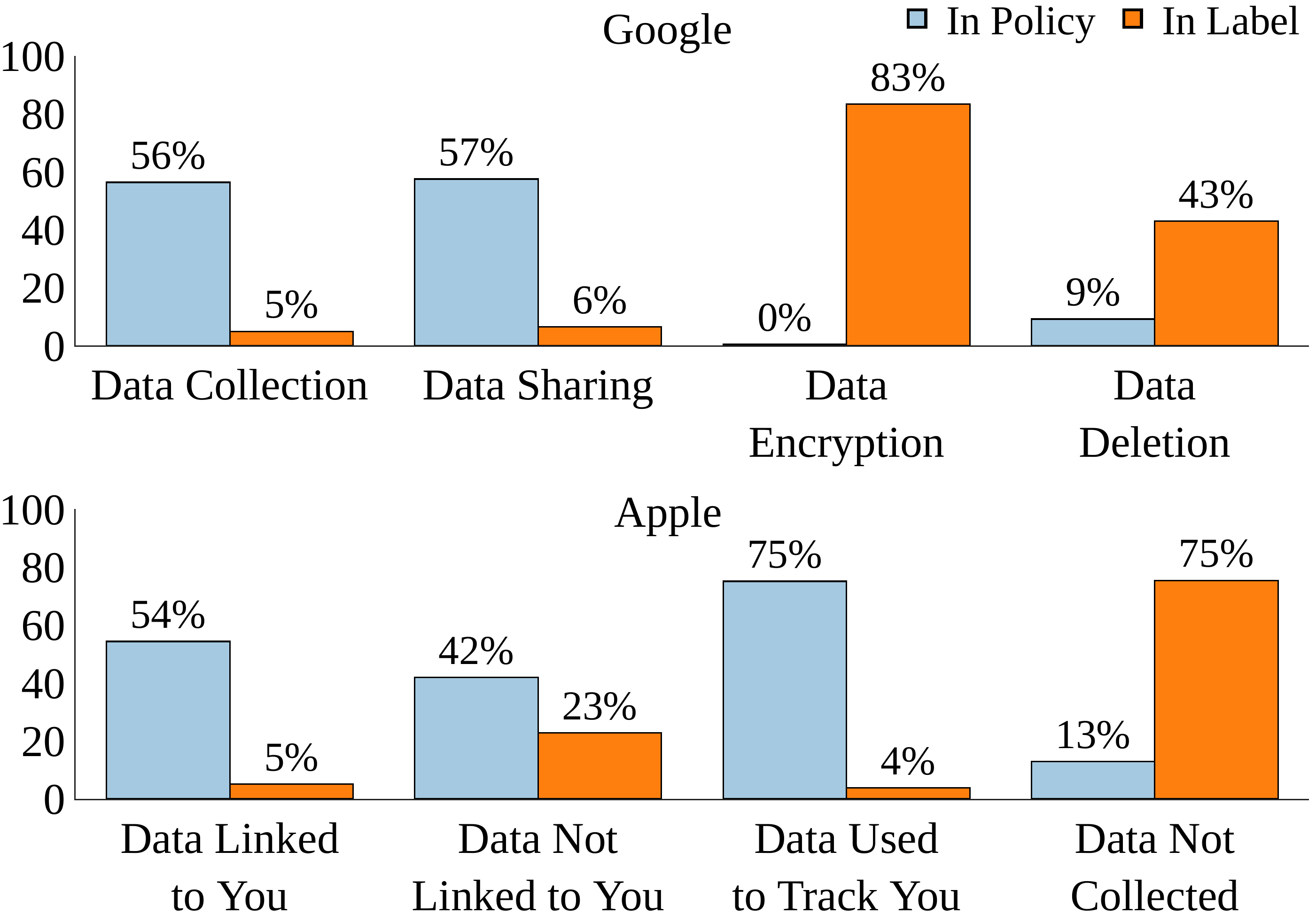}
  \caption{Inconsistencies between privacy policies and DSS and APL. The normalization is based on the total number of apps with privacy labels and classified policies.}
  \label{fig:policy_comp}
\end{figure}

\cref{fig:policy_comp} shows the \textit{In Policy} and \textit{In Label} inconsistency for high level practices for apps on the play store. We note that only 5\% (6\%) of the apps with DSS that collect (share) data have \textit{In Label} inconsistency with their privacy policies. 
% \rishabh{@Asmit: Add one example here, possibly from one of the emails that we got.} 
We also find that \textit{In Policy} inconsistencies for these categories are more than 55\%, but as discussed above, these could be due to privacy policy covering multiple apps and websites. To understand the extent to which this happens due to multiple apps, we analyze DSS for apps from the same developers. There are 15,380 developers who have 3 or more apps. These developers have an average of 13 apps and a median of 7 apps per developer. We find that 68\%(10,420) of these developers have duplicate data safety section for their apps. 
%The mean percentage of duplicate data safety section among all of these developers is 68.2\%, with a standard deviation of 22\% and a median of 75\%. 
For example, the app developer \textit{Premium Software} has over 9 apps across 6 genres but with only 2 unique DSS. 
% This indicates that \textit{In Policy} inconsistencies for developers having multiple apps are either coming from practices of websites or  are legitimate inconsistencies. 
This also highlights that developers might be duplicating their DSS across their apps, even though the apps can span multiple genres and have different features. 
% \rishabh{Maybe we need an example here? Or remove this part altogether?}

Analyzing the inconsistencies for \textit{Data Encryption} and \textit{Data Deletion}, we find that the majority of apps declare them in their privacy labels but there is no mention of such practices in their privacy policies. For example, \textit{Snapchat} mentions in their DSS that data is encrypted in transit but no corresponding practice is present in their privacy policy. Similarly, \textit{Kik — Messaging \& Chat App} state in their DSS that \textit{Data can’t be deleted} yet their privacy policy states that users can ask them to delete their information. It is worth noting that from a privacy and regulation standpoint, these two practices are extremely important. \textit{Data Deletion} option gives the users the right to either delete their data or ensure that it stays in anonymized form, which has roots in several regulations such as the GDPR~\cite{linden2018privacy} and the CCPA~\cite{ccpa}. \textit{Data Encryption} on the other hand, is crucial to prevent data snooping attacks which aim to get unlawful access to the data while the data is in transit.

% From analyzing the data gathered on apps from the Google Play Store, as shown in the top plot in \cref{fig:policy_comp}, we can see that for \textit{Data Collection} and \textit{Data Sharing}, over 50\% of apps have some type of inconsistency between their privacy policies and labels. Moreover, we note that for a majority of these apps their privacy policy mentions \textit{First Party} or \textit{Third Party Data Collection}, yet their privacy labels do not reflect that. For example, according to the DSS of the app, \textit{ShareMe: File sharing}\footnote{\url{https://play.google.com/store/apps/details?id=com.xiaomi.midrop}}, no data is being collected but their privacy policy does state that they can collect data. Interestingly this app also shares data with third parties but doesn't collect any data, which is non-intuitive but can happen when third party libraries are used. 

Analyzing practices at the category level, we find that there are significant inconsistencies between Privacy Labels and policies for data sharing and data collection. Specifically, we find that for data sharing, 89\% of the apps have inconsistent (\textit{In Label}) Privacy Labels for \textit{Location}, 82\% of the apps for\textit{Device IDs}, and 74\% of the apps for \textit{Health and Fitness}. For example, \textit{Myntra - Fashion Shopping App} states that they collect location, health info, contact list, and much more, yet its privacy policy doesn't mention the collection or sharing of such data types. Similarly, \textit{Tripadvisor: Plan \& Book Trips} states that they collect and share location data yet there is no mention of such practices in their privacy policy. This suggests that developers report more precise data-sharing practices in Privacy Labels, and can inform users allowing them to make better choices.

\subsection{Apple Privacy Label}
Next, we analyze the apps on the App store and compare privacy practices mentioned in the apps' privacy policies and their Apple Privacy Labels. The bottom plot in \cref{fig:policy_comp} shows inconsistencies for the high-level categories present in APL. We find that for \textit{Data Linked to You} and \textit{Data Used to Track You}, the \textit{In Label} inconsistency is 5\% and 4\% respectively. As the other two of the high-level practices, \textit{Data Not Linked To You} and \textit{Data Not Collected} are negations, we consider the \textit{In Policy} inconsistency (see \cref{sec:policy_inconsistency}). We find that 42\% of the apps have policies that state that they do not link data whereas the privacy label indicates otherwise. Furthermore, 13\% of the apps have policies that do not have data collection or sharing, but the privacy label indicates otherwise. For example. ~\textit{Superior Vision} app on App Store states that they collect \textit{Health \& Fitness} data yet their policy doesn't state that.

In \cref{fig:policy_comp}, we observe that \textit{In Policy} inconsistency for \textit{Data Used to Track you} is very high. This implies that privacy policies include tracking practices while privacy labels do not. This can potentially be due to the presence of segments related to cookies in the privacy policy, for example. ~\textit{Netflix}'s privacy policy talks about using cookies to track users on their site but not about tracking via their app.

We next examine the consistency at the data category level for \textit{Data Linked to You} and find that, similar to DSS, \textit{Location} (39\%), \textit{Identifiers (51\%)} and \textit{Health and Fitness} (60\%) had the largest \textit{In Label} inconsistencies. For \textit{Data Not Linked to You}, we find inconsistencies primarily in the same data categories. For example, ~\textit{Jetpack Joyride} states that they collect Location data but their privacy policy states that they collect Location based on IP Address and not the GPS location.

% A similar analysis of Apple App Store apps, show that on average 30\% of apps have inconsistencies between what is reported by the APL and what their privacy policy states. Similar to what we observed in google apps, a considerable percentage of apps's policy mention that they collect data, anonymously or not, and use the data for tracking but their APLs fail to reflect these practices. \red{A notable example would be \textit{Netflix}\footnote{\url{https://apps.apple.com/us/app/netflix/id363590051}} which do not mention of any data being shared, their policy does share data}
% \todo[inline]{Netflix has Third Party Ads but not Data Used to Track You ---!?!??!}

\subsection{Takeaways}
In this section, we find that at least 40\% of the apps with DSS, and APL are inconsistent with their privacy policy. Additionally, we note that DSSs contain more information about \textit{Security practices} than privacy policies, and thus can provide useful information to the users. We also note that sensitive datatypes such as \textit{Location}, \textit{Identifiers} and \textit{Health and Fitness} had the largest \textit{In Label} inconsistencies, indicating that the developers disclose collection/sharing of these fine-grained datatypes in the privacy labels. 
\section{Data Practices Across Platforms}
\label{sec:consistency_cross}
Next, we compare the self-reported privacy practices in Privacy Labels of cross-listed apps across the Google play store and Apple app store using the cross-listed dataset described in \cref{sec:cross_apps}. 
% As DSS and APL use different terminology for privacy practices, we first create a common mapping for privacy practices. We then compare the practices reported in privacy labels.
%In particular, we first identify the apps that are present in both the platforms, extract their privacy nutrition labels and investigate whether the same app (and developers) report different privacy practices across different platforms. We first begin by detailing the methodology for identifying cross-listed apps and then perform the comparison between self reported privacy practices at scale. 

\begin{figure*}
\hspace*{-2.2cm} 
  \centering
  \includegraphics[width=1.3\columnwidth]{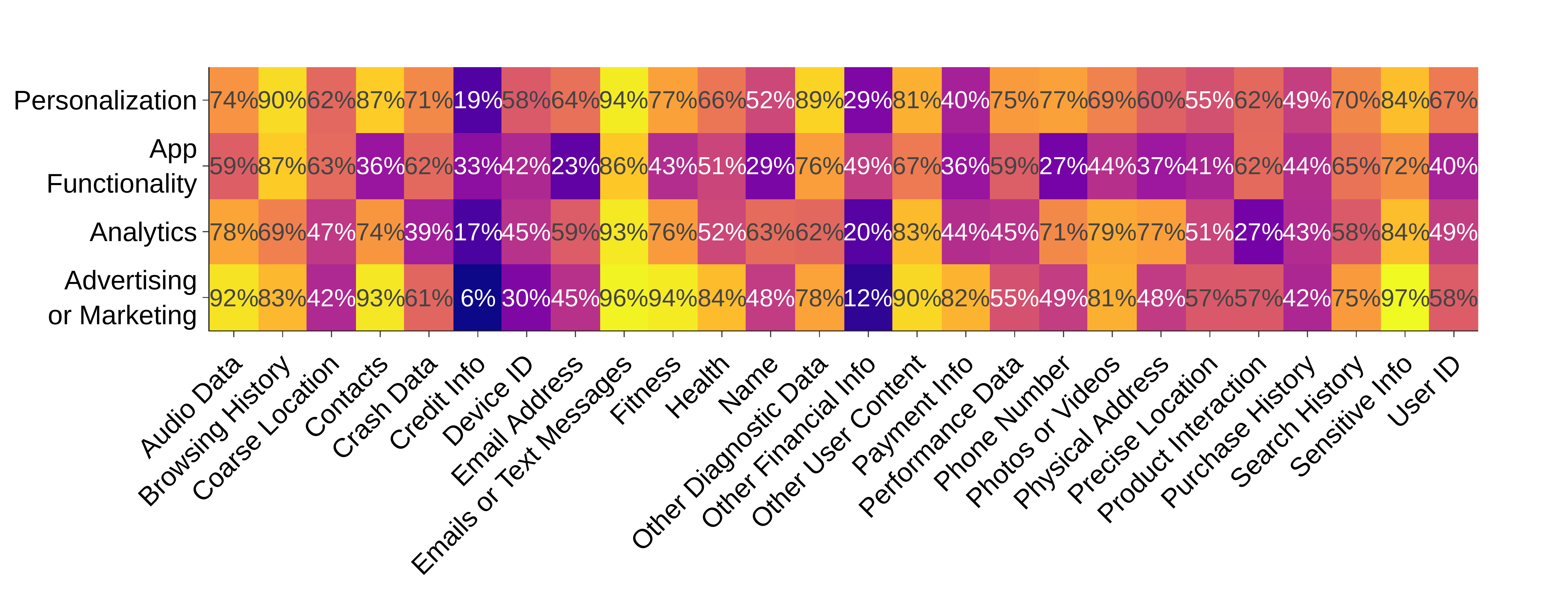}
  \caption{Normalized Heatmap showing the inconsistencies in the datatype-purpose pair. Normalization is done for each cell block in the heatmap, \textit{i.e.}, for each datatype-purpose pair, we normalize with the total number of apps which have that datatype-purpose pair.}
  \label{fig:heatmap}
\end{figure*}

\subsection{Mapping DSS categories to APL categories}
%Privacy Nutrition labels are designed to provide information about the data collection, sharing and usage by a given app. 
As previously discussed in Sec.~\ref{sec:background}, the privacy labels for android and iOS platforms cover different aspects of data practices. APL emphasizes on tracking and linkability of the collected data without distinguishing between collected and shared data, while DSS focuses on security practices and whether the data is collected or shared with a third party. 
%For example, in Apple privacy label, data used for tracking is highlighted separately whereas in data safety sections, security practices such as data encryption in transit and option to delete data are mentioned. 
However, despite covering disjoint high-level practices, the lower-level attributes in the privacy labels - namely the \texttt{datatype} and \texttt{purpose} - have a large overlap. Thus, to compare the disclosure practices of app developers across platforms, we first find the common datatypes and purposes in the two labels, and then compare 1) the datatypes, and 2) datatype-purpose pairs. 
% For the latter, we take a union of  all the common datatype-purpose pairs present in the APL and compare them with the common datatype-purpose pairs present in DSS. 

% \begin{table}[t]
% \footnotesize
% \centering
% \begin{tabular}{>{\arraybackslash}l>{\arraybackslash}l>{\arraybackslash}l}
% \toprule
% \multicolumn{1}{c}{\textbf{Google Purposes}} & $\rightarrow$      & \multicolumn{1}{c}{\textbf{Apple Purposes}} \\ \midrule
% Approximate Location  &$\rightarrow$  & Coarse Location  \\ \midrule
% Race and Ethnicity   & $\rightarrow$ & Sensitive Info\\ \midrule
% Sexual Orientation  & $\rightarrow$ & Sensitive Info \\ \midrule
% d &  &d \\ \midrule
% Personalization  & $\rightarrow$ & Personalization \\ \midrule
% Account Management       & $\rightarrow$ & N/A  \\  \midrule
% Developer Communication  &  & \\
% \bottomrule
% \end{tabular}}
% \caption{Table showing the common mapping from Data Safety Card and Apple Privacy Label}
% \end{table}

It is worth noting that the datatype and purpose tags used in the two labels can be used to denote different concepts. For example, in APL, \texttt{App functionality} also includes fraud prevention and implementing security measures, whereas the data safety section has separate tags for app functionality and fraud prevention and security measures. For the purposes of this analysis, we combine these two purposes into \textit{App functionality} to create a common map. Since APL does not have any tags for \textit{Account Management} and \textit{Dev. Communicaitons}, we removed them from DSS for this comparison. After taking the intersection of the available datatypes and purposes, we end up with 4 purpose categories and 26 datatypes. A complete mapping for classes from DSS to APL is shown in \cref{appendix:mapping}.
%There are also data types which are not common in the labels such as Files and Docs, Calendar etc. For our comparison, we remove these data types, along with the ambiguous data types such as \textit{Other info} in personal information and \textit{Other actions} (in App interactions). ~\cref{table:common_map1} and ~\cref{table:common_map2} provides a complete mapping that we have used to compare.

\subsection{Findings}
We compare the self-reported privacy practices of the 100K apps that are cross-listed on both the platforms and have privacy labels. Specifically, we ask the following questions: a) How does the high-level practice of data collection compare between the two labels? and b) Is the purpose for using datatypes consistent between the two labels?

% \begin{itemize}[leftmargin=*, align=parleft, labelsep=4mm]
%     \item[\textbf{Q1}] How does the high level practice of data collection compare between the two labels?
%     \item[\textbf{Q2}] When the apps uses datatypes, is the purpose for using datatypes consistent between the two labels?
% \end{itemize}

\subsubsection{Comparison of Data Collection}
 To compare how many apps do not collect data, we rely on the \textit{Data Not Collected} tag for the iOS platform and \textit{Data Shared} and \textit{Data Collected} tags for the android platform. 
 % Note that here we use both data shared and data collected tags for android platform because iOS platform denotes both the practices with 
 % does not make a distinction between the two, and the \textit{Data Not Collected} refers to both collection and sharing. 
 % We  emphasize here that due to the reason mentioned above, it is not possible to compare the data sharing practices between the two labels. 
We find that a total of \nnumber{22K(~22\%)} apps report different data collection practices on the two platforms. Of these apps, \nnumber{42\%} of the apps report collecting data on android while \nnumber{58\%} of the apps report collecting data on the iOS platform. Examining these apps further, we find that \nnumber{18\%} of these apps have more than 100k downloads, and \nnumber{5\%} has over 1M downloads indicating that even popular apps have this inconsistency. For example, ~\textit{KineMaster - Video Editor} a video editing app with over 400M+ downloads on Google Play Store states that they do not collect any data in the Play store but states in App Store that they do collect sensitive data such as \textit{Location} and \textit{Identifiers}.

The inconsistency in self-reported data collection practices as indicated by the inaccuracies in privacy labels undermines the credibility of the Privacy Label framework. This poses a significant concern for users, as they may base their decisions on inaccurate information, thereby increasing their privacy and security risks.

% The inconsistency suggests that at least one of the privacy labels is inaccurate which is concerning as this indicates that developers are not consistent in self reporting their data collection practices, which in turn erodes on the credibility of the Privacy Label framework. If the privacy labels are inaccurate, then privacy-concious users might be at higher privacy and security risk as they may make decisions about which app to use based on false information.

\subsubsection{Comparison of Fine-grained Practice}
% Next, we investigate whether the apps mention the same purpose for using datatypes in the privacy labels across the platforms. To perform this analysis, we only consider the apps where both the labels report collecting data, resulting in a set of \nnumber{yy} apps.
%removed the 50K apps identified above from this analysis as our objective is to analyze fine-grained inconsistencies across the two privacy labels when both the apps are collecting data. 

%Our assumption here is that same app will have the same features across the platforms and therefore will request same datatypes on both the platforms for same purpose. We note that neither of the platforms disallow developers from collecting or using a particular data type. The underlying operating systems might provide more controls to the user, but the developers are not prohibited from accessing any resources. Further, we note that we removed the 50K apps identified above from this analysis as our objective is to understand 

\begin{figure}[t]
  \centering
  \includegraphics[width=\columnwidth]{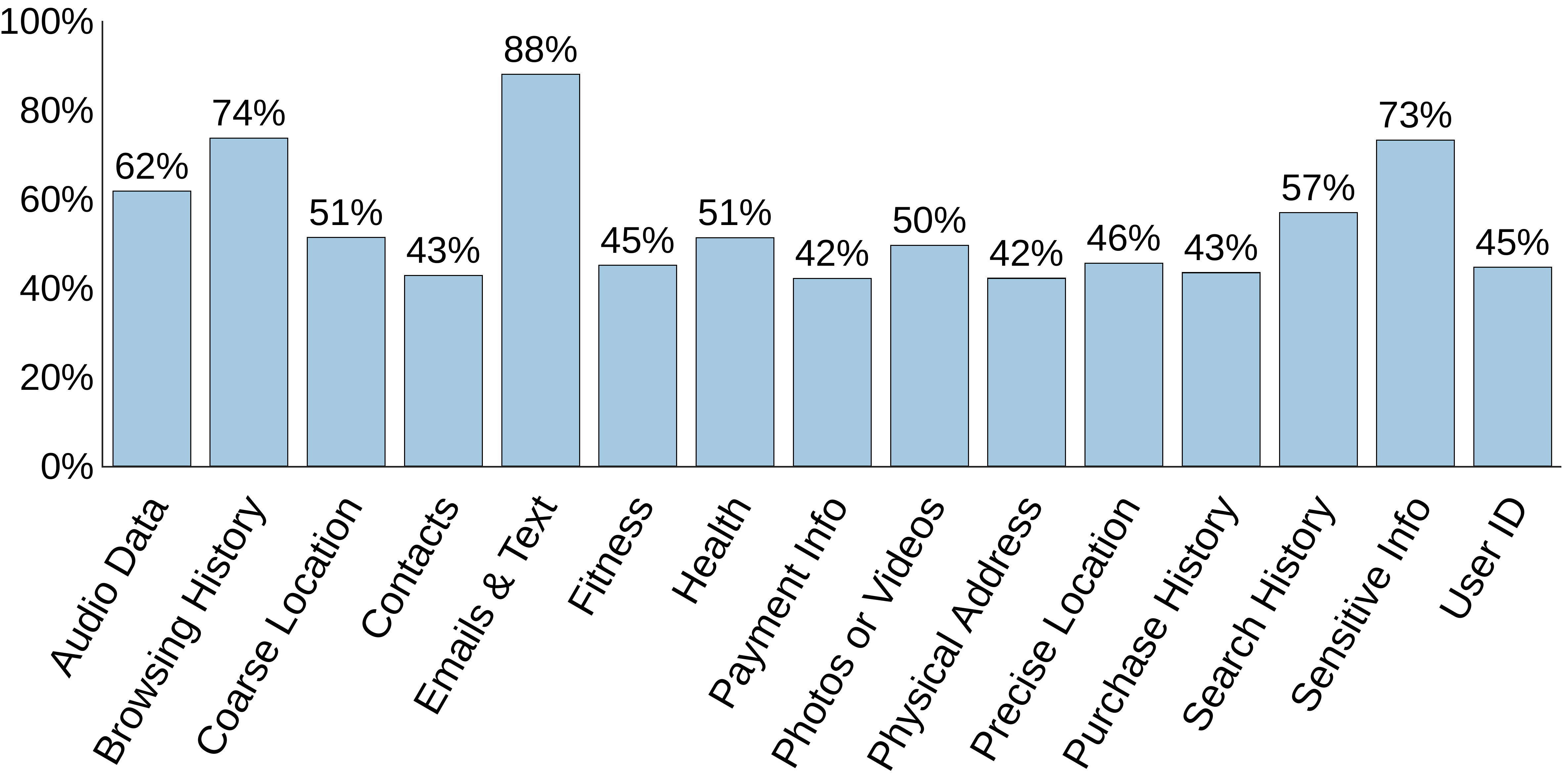}
  \caption{Distribution of inconsistent apps with datatypes. Each datatype is normalized with the number of apps using that particular datatype on either platform. Note that we have omitted some datatypes here for brevity. The full distribution can be found in the ~\cref{fig:inconsistency_app_no_thresh} in the Appendix}.
  \label{fig:inconsistency}
\end{figure}

We compare fine-grained practices along two dimensions: 1) DataType where we check whether the privacy labels report collecting/sharing the same datatypes; and 2) DataType-Purpose pairs where we compare the common datatype-purpose pair in the two labels. If there is at least one datatype-purpose pair that is not present in both sets, we treat the app as inconsistent for that datatype-purpose pair. We also tag instances as inconsistent where the datatype-purpose pair is present in one of the labels and is missing from the other. For example, if the DSS of an app has \textit{(Location - Personalization)}, while the APL has \textit{(Location - App Functionality)}, then we treat the app as inconsistent. Similarly, if the DSS of an app has \textit{(Location - Personalization)}, while the APL has \textit{(Location - App Functionality)} and \textit{(Location - Personalization)}, we still treat the app as inconsistent as the tags \textit{(Location - App Functionality)} is not common in both. 

Across the cross-listed apps that have privacy labels, we find that at least \nnumber{60\%} of the apps have at least one inconsistency. For example, in \textit{Tiktok}, DSS states that they collect the contact list of the users for `Advertising and Marketing' purposes, but APL states that the app does not collect a contact list. 
\cref{fig:inconsistency} shows the inconsistency in datatypes across the two platforms. We find that \textit{Sensitive Information}, \textit{Browsing History}, and \textit{Emails or Text Messages} have the highest inconsistencies across the two platforms.  From Fig.~\ref{fig:inconsistency}, we observe that DeviceID and Product Interactions are the data counts with the highest inconsistencies. We also observe that \textit{Precise Location} and \textit{Coarse Location} is inconsistent with \textit{Advertising} implying that at least in one of the labels, location is used for advertising, raising privacy concerns for the users.

% Analyzing inconsistencies in purpose \rishabh{Add this when the plot is ready}. 

% We find that at least \red{70\%} of the apps have at least one inconsistency between the two privacy labels. Fig.~\ref{fig:inconsistency} shows the distribution of inconsistent apps with data types and with purpose. We find that \textit{App functionality} is the most common purpose for which inconsistencies arise. For Datatypes \textit{DeviceID} is the most common data type for which there are inconsistencies. For example, <AAP A>, a popular app with XX downloads has ...
% Only looking at the popular apps, we see a similar pattern with App Functionality and DeviceID being the most common purpose and datatype respectively, with inconsistencies. 

To analyze datatype-purpose inconsistencies, we show the normalized heatmap of inconsistent apps in \cref{fig:heatmap}. Normalization is done for each cell block in the heatmap, \textit{i.e.}, for each datatype-purpose pair, we normalize with the total number of apps that have that datatype-purpose pair. 
%Thus, each cell block in \cref{fig:heatmap} denotes the fraction of apps in which the corresponding datatype-purpose pair is consistent. 
We find that \textit{Fitness} and \textit{Sensitive Information} when used for \textit{Advertising or Marketing} are frequently inconsistent. The plot shows that even though \textit{Sensitive Information} and \textit{Fitness} data are not collected very often (Fig.~\ref{fig:inconsistency}), when they are collected, they are often inconsistent in privacy labels across two platforms. On the other hand, \textit{Credit Information} and \textit{Financial Information} have the least number of inconsistencies, which is encouraging considering the sensitive nature of this information.

\subsection{Takeaway}
In this section, we analyzed the consistency of privacy labels for the same apps across the two platforms. We find that 60\% of the cross-listed apps had at least one inconsistency between APL and DSS. We further find that inconsistencies are highest for \textit{Sensitive Information}, \textit{Browsing History}, and  \textit{Emails or Text Messages} datatypes. Through a detailed analysis of datatype-purpose inconsistencies, we find that \textit{Emails and Text Messages} when used for Advertising results in inconsistencies 96\% of the time, indicating a concerning problem with disclosure of practices in privacy labels.

\section{Discussion}
In this paper, we investigated the consistency of privacy labels with privacy policies and labels on other platforms. Our findings suggest that there is a significant degree of inconsistency in privacy labels. Overall, there is a need for greater consistency in the way that privacy practices are disclosed to users, both within and between platforms. In this section, we discuss the implications of our findings and suggest potential solutions for improving the transparency and consistency of privacy practices. We also discuss the limitations of our study.\smallskip

\noindent\textbf{Comparison between the two labels.} We analyzed both the Data Safety Sections and the Apple Privacy Labels and find that the two labels cover different aspects of data practices. While both labels provide information about the types of data that apps collect, Apple's privacy label does not distinguish between data collection and data sharing. Apple's privacy label is more explicit about certain aspects of data practices, such as linkability, third-party advertising, and tracking, whereas data safety sections lack these details, but does inform the users about the safety of their data (\textit{Data Encryption}) and the choices that they have with developers (\textit{Data Deletion Option}). These practices may be of particular interest to the users in light of the GDPR~\cite{linden2018privacy}, which requires companies to provide a clear and explicit purpose for the collection and use of personal data. The regulations like the GDPR and the CCPA also provide the right to delete the data to the users, which is covered in Data safety forms but not in the Apple privacy labels.

The comparison between the two labels highlights the importance of considering multiple sources of information when evaluating the data practices of apps. By combining the information provided by both labels, users can make more informed decisions about their privacy and the apps they choose to use.\smallskip

\noindent\textbf{Usability of Privacy Policies For Apps.} Privacy policies have been used as a default framework for notice and choice to users. Our analysis reveals that many developers have several products, including websites, Internet of Things (IoT) devices, and applications. However, it is common for these products from the same developer to have the same privacy policy, even if they collect data in different ways. This provides inaccurate information, as the privacy policy itself may not accurately convey the privacy practices of a specific product. This can be addressed by having separate privacy policies for each product or by clearly identifying the specific practices that apply to each product within a single privacy policy. Failing to do so may lead to misunderstandings and mistrust among users, and may also violate privacy regulations.\smallskip

\noindent\textbf{Inconsistencies in disclosed practices across platforms.} Our findings indicate that there are inconsistencies between the privacy labels in the Apple Privacy Labels and the Google Data Safety Sections for the same apps. One possible reason for these inconsistencies is the confusing framework for privacy labels. While previous research~\cite{li2022understanding} has shown that privacy labels are useful for both developers and users, it also highlighted that filling privacy labels is perceived as challenging extra work. On top of that, developers are also unclear about definitions which can result in confusion and ultimately, inaccurate privacy labels. This confusion can be compounded by the fact that different platforms may use different terminology to describe similar practices. For example, in Apple's privacy label, the term \textit{tracking} is used when data collected is linked with third-party data for advertising purposes or when data is shared with a third party, which can be confusing to the developers, even when they are asked to pay close attention~\cite{li2022understanding}.

Another possible reason for the inconsistencies we observed is the casual attitude of some developers toward disclosing their data practices. Some developers may not fully understand the data practices of their own apps, or may not prioritize accurately disclosing this information to users. Finally, the platforms lack consistency checks to ensure that the information provided in the privacy labels is accurate. Without these checks, it is possible for developers to provide misleading or incomplete information about their data practices, just to meet the requirements.

We note that these inconsistencies can have serious consequences for users, as they may be confused about the privacy practices of the apps they use. If the practices disclosed in the privacy labels are inaccurate, it can reduce the efficacy of these labels as a tool for helping users make informed decisions about their privacy. Even worse, it could induce a false sense of security in users, who may assume that their data is being handled in a certain way when it is not.\smallskip

\noindent\textbf{Usability of Privacy Labels.} Even though our analysis finds inconsistencies between privacy labels and privacy practices, evidence suggests that privacy labels generally carry more specific information about the practices. They include information about the types of data that an app collects, how the data is used, and whether it is shared with third parties. This information can be very useful for users who are concerned about their privacy and want to ensure that they are only using apps that respect their personal data.

However, the accuracy of privacy labels is not guaranteed. While developers are required to disclose their data practices in order to obtain a privacy label, there is no guarantee that the information they provide is accurate or complete. As such, it is important for platforms to recognize that developers may not always be honest about their data practices. Therefore, it is necessary to have systems in place to verify the accuracy of privacy labels and to hold developers accountable for any discrepancies. This is particularly important because the false labels can create a false sense of security among the users. 

One potential model for regulating privacy labels is a system similar to the one used for food nutrition labels, which are regulated by the Food and Drug Administration (FDA). A regulatory body could be established to oversee privacy labels and ensure that they are accurate and consistent. This could help to build trust among users and encourage developers to be more transparent about their data practices.\smallskip

\noindent\textbf{Limitations.} Extracting privacy practices using automated analysis comes with several limitations. First, the framework used here treats privacy policies as segmented text, missing out of relations between different segments. This can potentially result in internal contradictions, as shown in ~\cite{andow2019policylint}. Second, the classifiers used to extract privacy practices can introduce errors, which can then propagate through the pipeline and induce uncertainty in the inconsistency rates. We do however note that the error rate of our classifiers is significantly less than the inconsistency rate obtained, indicating that the results presented in the paper are valid. 

% \begin{itemize}
%     \item Inconsistency via data flow analysis
%     \item Limitations of Automated Policy analysis, specially for practices which can long range relations
% \end{itemize}

% \begin{itemize}
%     \item Comparison between the two labels: Talk about how the two labels combined can more information.\\
%     Subpoint: The two labels are covering very different aspects data practices. Apple does not distinguish between colelction and sharing, whereas google is not explicit about linkability, third party advertising, tracking etc. Connect these to the GDPR
%     \item possible reasons for the discrepancy, both for privacy policies and for labels
%     \item Talk about privacy policies not being adequate for conveying information about the practices and they cover vastly different informations. This is sort of a motivation for using privacy labels, but the information provided there is not reliable as well. 
%     \item Limitations: talk about automated privacy policy analysis.
%     \item Future direction: Use privacy labels to automatically edit privacy policies and generate a more comprehensive/readable privacy policies
% \end{itemize}
\section{Conclusion}

In conclusion, our large-scale measurements of Privacy Labels have provided valuable insights into the privacy practices of apps. By analyzing Data Safety Sections for 2.5M apps and Apple Privacy Labels for 1.38M apps, we provided a comprehensive picture of the privacy practices of the applications. On one hand, privacy labels provide users with more specific information about the data practices of apps than traditional privacy policies. However, our analysis showed that there is often a discrepancy between the information disclosed in privacy labels and the information contained in privacy policies. This can be confusing for users and may make it difficult for them to make informed decisions about which apps to use based on their privacy concerns. Furthermore, our comparison of Privacy Labels for cross-listed apps in the Play store and Apple store showed differences in the practices disclosed, indicating that developers are not consistently disclosing the same information on different platforms. Overall, these findings highlight the importance of carefully reviewing Privacy Labels and other sources of information when evaluating the privacy practices of apps. They also suggest that there is a need for improved transparency and accountability in the app industry, as developers may not always be accurately disclosing their data collection and use practices. Having a more transparent system will allow the consumers to be aware of the data collection and use practices of the apps and make informed decisions about their privacy. 
    
% \newpage
% \bibliographystyle{IEEEtranS}
\bibliographystyle{abbrv}
\bibliography{ref.bib}

\appendix

\section{Privacy Policy Analysis}
\subsection{Privacy Policy Taxonomy}
\label{appendix:taxonomy}

\begin{figure*}
\hspace*{-2.2cm} 
  \centering
  \includegraphics[width=1.25\columnwidth]{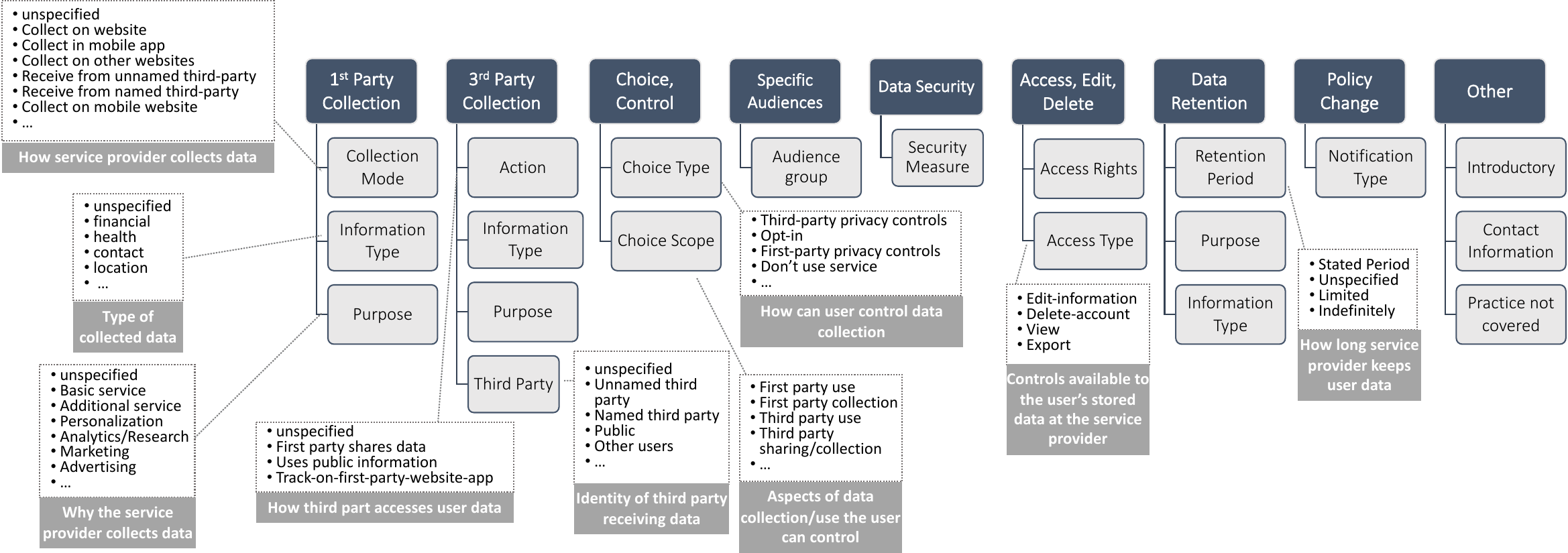}
  \caption{The privacy policy taxonomy by Wilson et al. ~\cite{wilson2016creation}}
  \label{fig:opp_taxonomy}
\end{figure*}

\textbf{Limitations of the OPP-115 Taxonomy} Figure.~\ref{fig:opp_taxonomy} shows the privacy taxonomy proposed by Wilson et al.~\cite{wilson2016creation}. The top-level defined high-level privacy categories whereas the lower level defined a set of privacy attributes that can take a particular set of values. Additionally, some examples of attribute-value pairs are shown such as Information Type and Purpose. Note that several lower-level attributes are shared across the high-level categories. 

Prior works~\cite{harkous2018polisis, srinath2020privacy} have used the OPP-115 taxonomy and the associated dataset to build machine-learning classifiers that tag segments of the policy with the labels from the taxonomy. However, there are two limitations to directly using the taxonomy (and existing frameworks such as Polisis~\cite{harkous2018polisis}) to compare privacy practices between privacy labels and privacy policies. First, the OPP-115 taxonomy was developed for privacy policies of websites, which is vastly different than the ecosystem of applications (both Android and iOS). In particular, the applications have access to sensitive data types, which are present in the privacy labels. This taxonomy, while having some overlap with the APL and DSS privacy labels, does not cover such app-specific data types. For example, \texttt{app activity}, a data category covering users' interactions within the application, is not covered in the taxonomy. Second, the OPP-115 dataset has limited annotations for the lower-level attributes that overlap with the private labels. For example, \textit{Encryption in Transit}, which is a separate practice covered in Data Safety Sections, only has less than 100 labeled instances in the OPP-115 dataset. 

\begin{figure}
  \centering
  \includegraphics[scale=0.6]{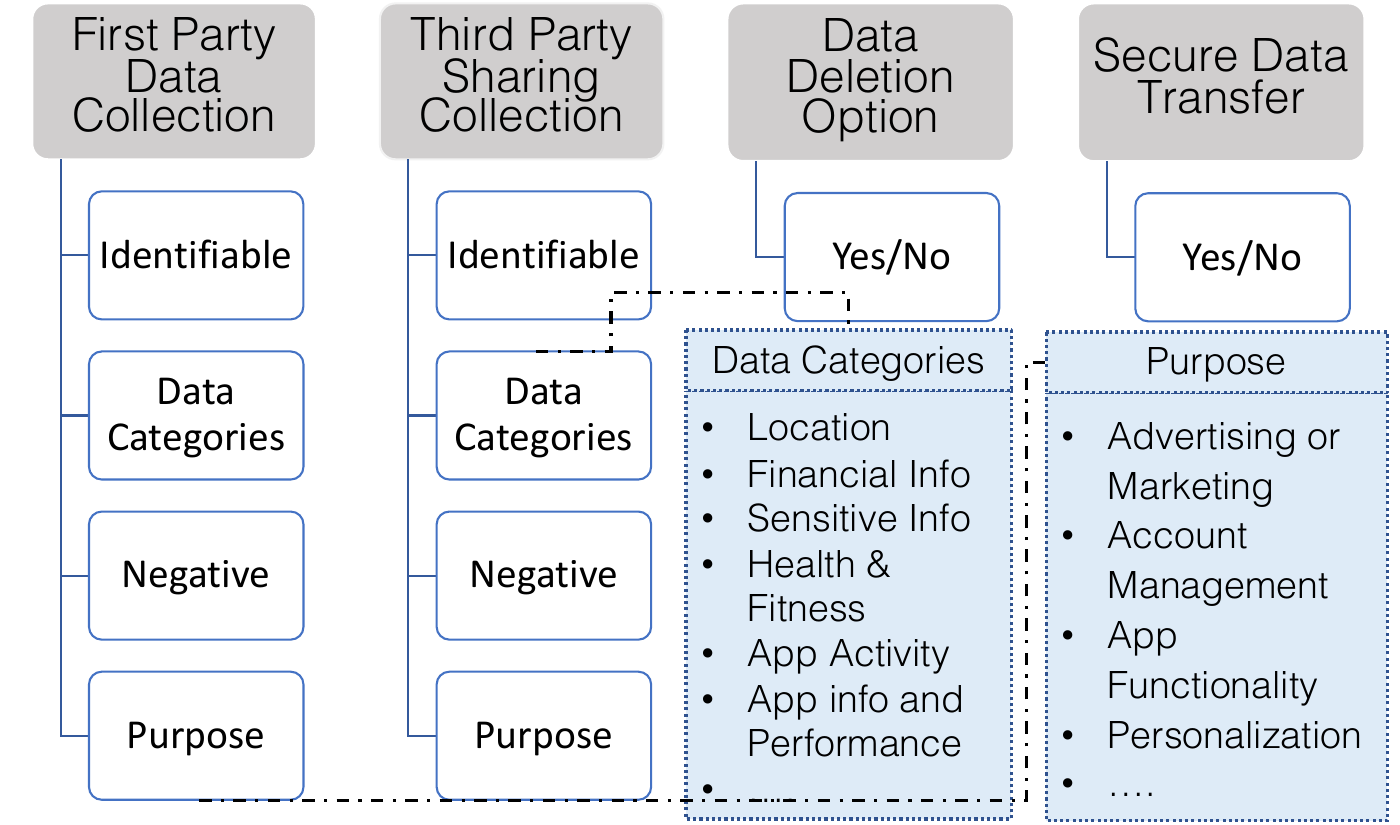}
  \caption{Privacy Label Taxonomy}
  \label{fig:privacy_taxonomy}
\end{figure}

We address these limitations by incorporating the missing labels to the existing OPP-115 taxonomy. We derive a \textit{Privacy Label Taxonomy} (\cref{fig:privacy_taxonomy}) as a union of a subset of the original OPP-115 taxonomy with the new labels from APL and DSS. To build the taxonomy, we first identify the categories from the taxonomy that are relevant to privacy labels, thus creating a subset of the original taxonomy. We then add the missing categories to get the new taxonomy. 

\medskip
\noindent \textbf{Identifying relevant categories from OPP-115} 
As discussed in Sec.~\ref{sec:background}, privacy labels consist of high-level privacy practices, data categories, and purposes for the use of data. The high-level categories \textit{First-party-data-collection} and \textit{Third-party-sharing-collection} from the OPP-115 taxonomy are relevant as they map directly to Data Safety Sections' \textit{Data Collection} and \textit{Data Sharing} privacy types. Further, APL covers the first-party collection and sharing practices implicitly through \textit{Data Linked to You} and \textit{Data Not Linked to You}. Similarly, the attribute level categories \textit{Purpose}, \textit{Data Type}, and \textit{Identifiable} are relevant.

For example, in Apple Privacy Label (APL), the \textit{Data Used to Track You} privacy type includes the data that is linked with third-party data for targeted advertising. It also includes cases when the data is shared with a data broker. Note here that linking can be done by both the app developers (by using data obtained from a third party) or by sharing the data with a third party. Thus, this privacy practice can be represented with \textit{Advertising or Marketing} purpose of \textit{First-party-collection-use} and \textit{Third-party-sharing}. At this stage, we drop the categories absent in the privacy labels. For example, \textit{Policy Change} is a high-level category in OPP-115 which is not present in the \textit{Privacy Label Taxonomy}. 

\medskip
\noindent \textbf{Adding New Categories} As indicated earlier, the OPP-115 taxonomy misses some of the lower-level data categories and purposes. We add these missing elements and adapt the OPP-115 taxonomy to \textit{Privacy Label Taxonomy}. 

Apart from the high level categories from the taxonomy, we also add two high level categories: \textit{Data Deletion Option} and \textit{Encryption in Transit}. Both the categories are part of \textit{Security Practices} privacy type from DSS. \textit{Data Deletion} corresponds to when the app ``Provides a way for you to request that your data be deleted, or automatically deletes or anonymizes your data within 90 days''. As there is no specific way to get this information from the taxonomy, we create a separate high level practice for Data Deletion. For \textit{Secure Data Transfer}, there is low level element in the taxonomy that covers the practice, however, since the other categories from the taxonomy in the hierarchy are not related, we add \textit{Secure Data Transfer} as a high level category. Also note that since there were less than 100 annotations for this category, we also perform additional annotations and increase the dataset size.

\begin{figure}[htbp]
  \centering
  \includegraphics[width=\columnwidth]{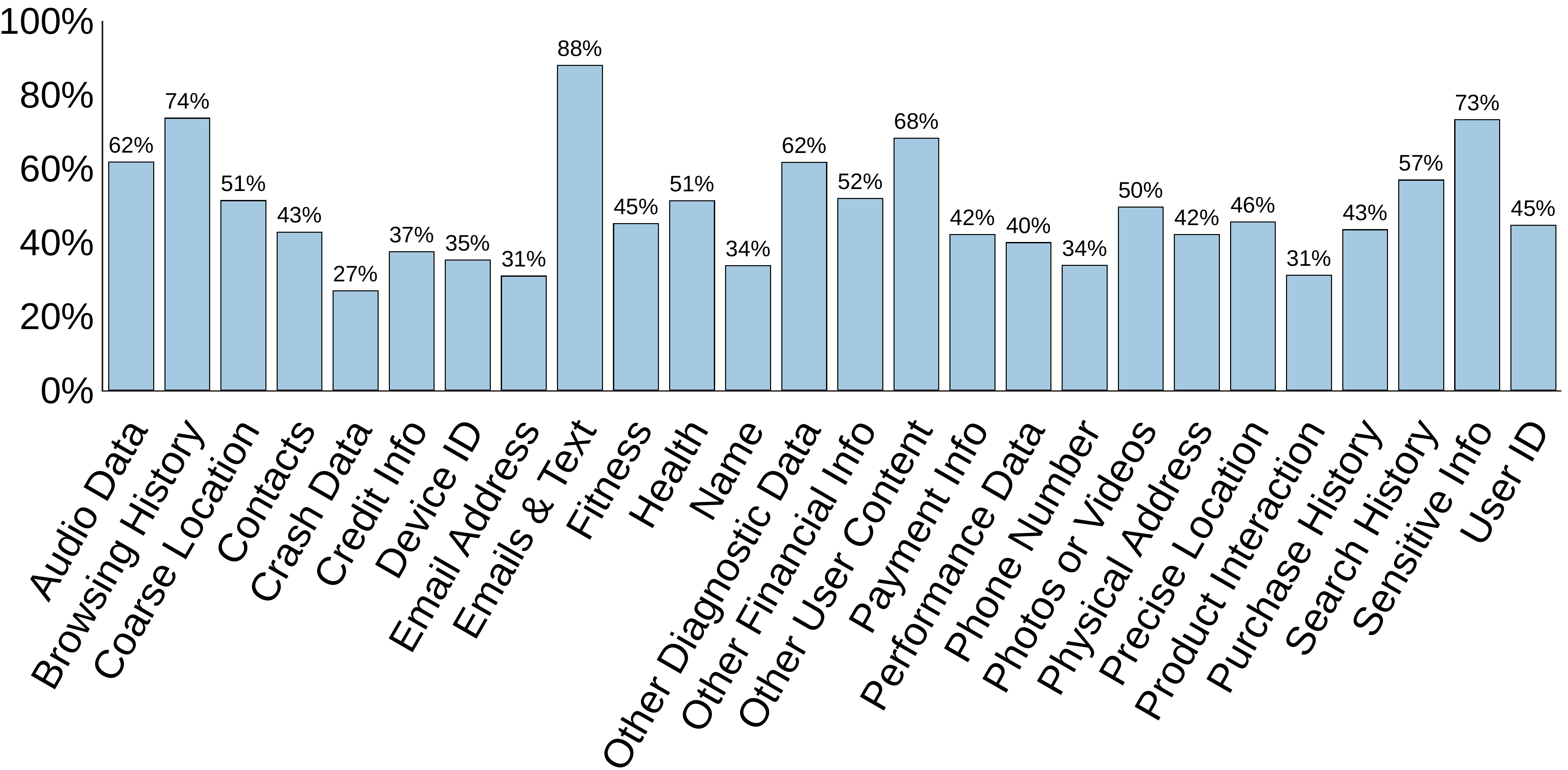}
  \caption{}
  \label{fig:inconsistency_app_no_thresh}
\end{figure}

% \begin{figure*}[h]
%   \centering
%   \includegraphics[width=2\columnwidth]{figures/good_pdfs/heatmap2_cropped.pdf}
%   \caption{}
%   \label{fig:inconsistency_app}
% \end{figure*}

\subsection{Annotation Setup}
\label{appendix:annotation_setup}

\noindent \textbf{Creating Annotation Set} For our \textit{Privacy Label Taxonomy}, we were able to have the data missing from OPP-115 for \red{13} elements. Curating the candidate set for missing categories is a major challenge due to label imbalance. To address this issue, we follow the approach used by Harkous et al.,~\cite{harkous2022hark} and use the task of \textit{Natural Language Inference} (NLI) to curate the candidate set. The NLI tasks consist of a hypothesis and a premise, and the objective is to determine if the hypothesis is true (\texttt{entailment}), false (\texttt{contradiction}) or undetermined (\texttt{neutral}) given the premise~\cite{maccartney2009natural}. For example, if the premise is: \textit{``Your data is safely and completely removed from our servers or retained only in anonymized form.''} and the hypothesis is \textit{``Data deletion is being discussed''}, then this instance will receive an entailment. On the other hand, if the hypothesis were \textit{``Policy change is being discussed''}, then the label would be neutral. This method of using NLI-based sampling to reduce the annotation effort has been shown to be effective by Harkous et al.~\cite{harkous2022hark}.

We start by creating a hypothesis for each of the missing categories that we have. For example, for \textit{Data Deletion Option}, we created two hypotheses: ``Data deletion is being discussed'' and ``Data Anonymization is being discussed''. For the NLI task, we used the T5-Large model checkpoint from \texttt{Huggingface}. This model is already trained on MultiNLI task~\cite{2020t5} which consists of a multi-genre dataset covering a large variety of domains.  Next, we run the NLI model and get weak labels for all the missing categories. Note that these are weak labels that are later manually annotated to create the training set.

\medskip
\noindent \textbf{Annotation Details} Using the NLI sampling approach, we curated a candidate set with 2000 segments for each of the missing categories. These segments are roughly balanced based on the weak labels assigned by the NLI model. For each class, we then randomly sample 500 segments to annotate. Two of the authors annotated the segments and created the training set.

The annotation was performed using the label studio framework~\cite{labstud}. The framework supports not only simple natural language processing tasks but also sophisticated labels such as taxonomies and sentence highlightings. The framework also supports active learning with the capability of integrating a backend machine learning classifier of one's taste in order to facilitate annotation.

\begin{figure}[htbp]
  \centering
  \includegraphics[width=\columnwidth]{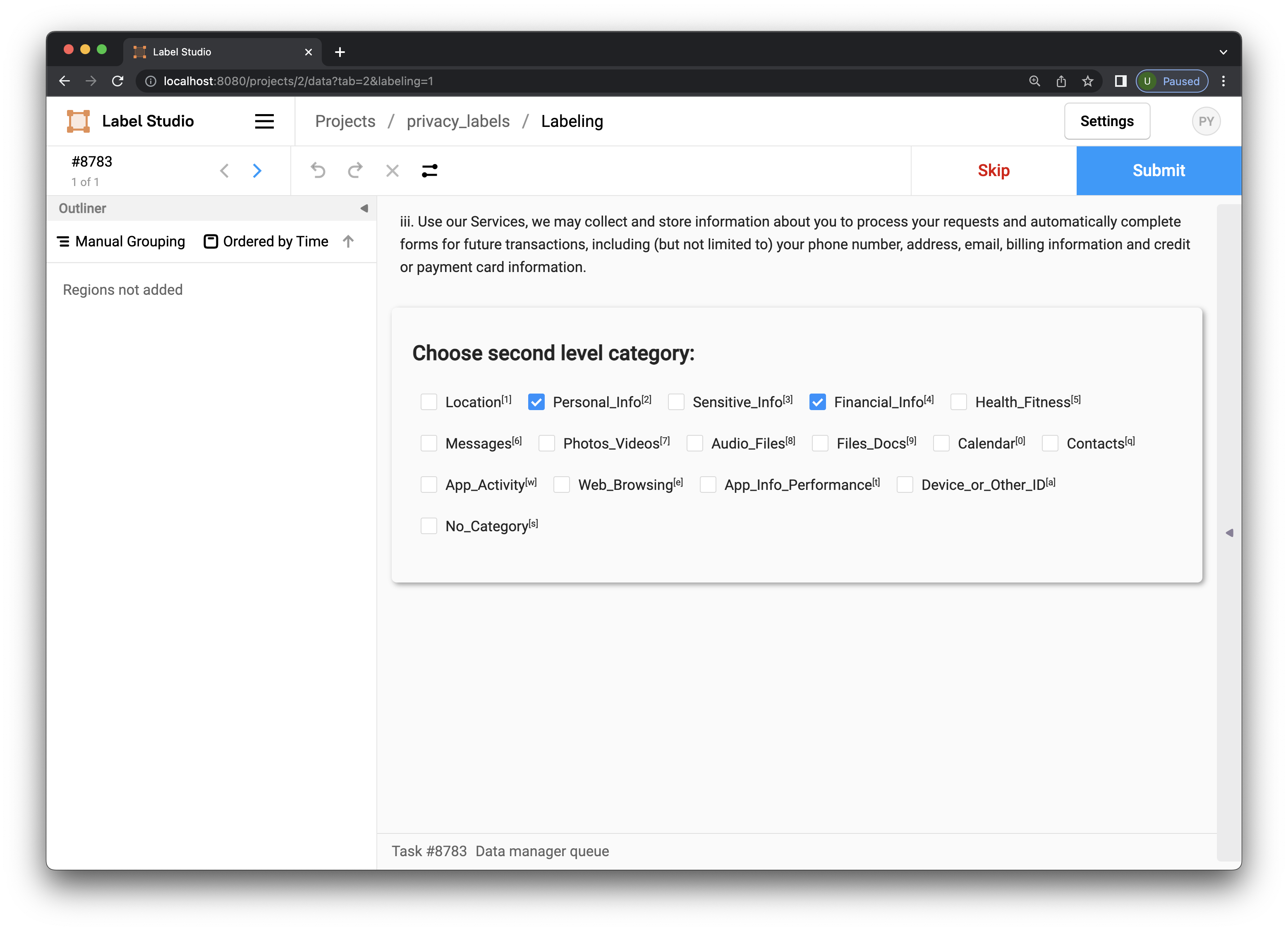}
  \caption{Label Studio Annotation Setup}
  \label{fig:anno_setup}
\end{figure}

The label studio server was deployed in an internal network, where the two authors simultaneously worked on annotating and creating the training set. Figure \ref{fig:anno_setup} shows the annotation setup used by the authors.

After the annotation step, the dataset was converted into a data frame consisting of the text column and a binary indicator column for each of the categories, to prepare for the training.

\subsection{Training Setup}
\label{appendix:training_setup}
Large language models like BERT~\cite{devlin-etal-2019-bert}, T5~\cite{2020t5} etc have shown remarkable performance using small training sets. Thus, for our purposes, we use the \texttt{distilbert-base-uncased}~\cite{sanh2019distilbert} model consisting of 67 million parameters. This model is the distilled version of BERT~\cite{sanh2019distilbert}. It has 40\% fewer parameters, can run 60\% faster and performs only slightly worse (\textasciitilde 5\%) than the original \texttt{bert-base-uncased} model on several natural language tasks. Additionally, we also perform domain adaptation by pre-training the DistilBert model on privacy policy text with the Masked Language Model (MLM) task. In particular, we pre-trained the model with the default hyperparameters, with a batch size of 256 for 24800 steps on a single NVIDIA A100 GPU.

We then use the new pre-trained model to train the category classifiers for the Privacy Label Taxonomy. We use a classification head on the model after adding a linear classification. For classification, the data annotated by the authors is split into two parts: testing (20\%) and training sets (80\%).

\section{Data Practices in Privacy Labels}
\label{appendix:landscape_details}
In this section, we look into the distribution of all the data types as shown in \cref{fig:inconsistency_app_no_thresh}. From this figure, we observe that the six categories--App Activity, App Info \& Performance, Device IDs, Financial Info, Location, Personal Info--are reported to be the most collected in Google, while the six categories--Contact Info, Diagnostics, Identifiers, Location, Purchases, Usage Data and User Content--are reported to be the most collected in Apple.

\begin{figure*}
\hspace*{-2.2cm} 
  \centering
  \includegraphics[width=1.25\columnwidth]{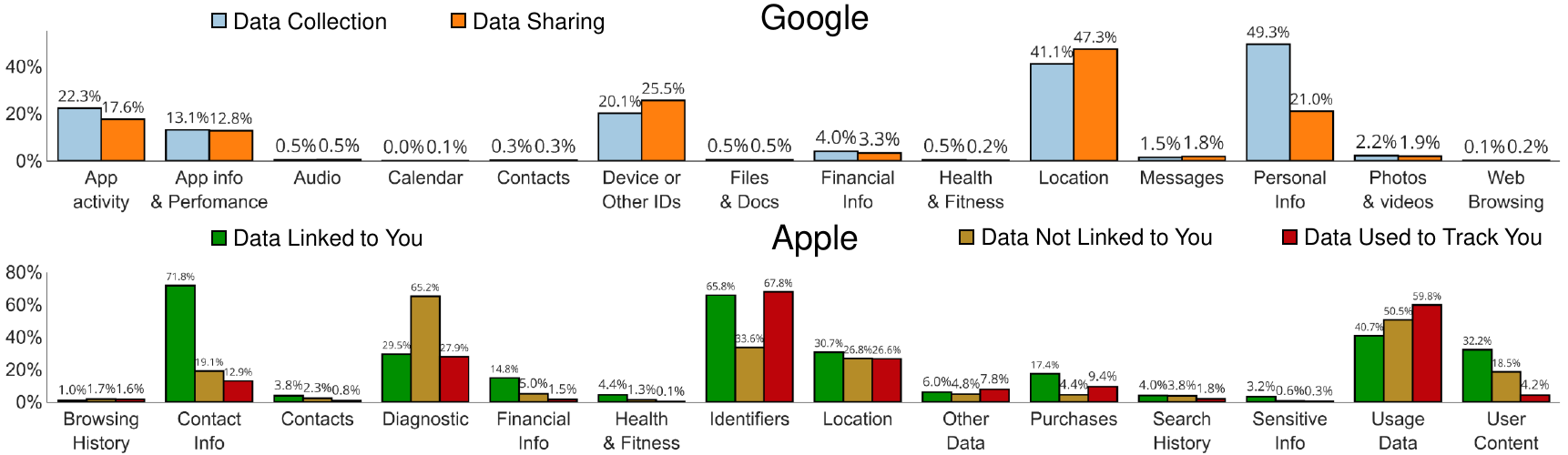}
  \caption{Distribution of data categories for high level
practices for apps in Play Store (top) and App store (bottom).}
  \label{fig:high_level_data_dist}
\end{figure*}

\section{Developer Study}
\label{app:dev-study}
For the Developer Study (\cref{sec:developer}) we sent emails to developers in 3 different categories: (A) apps stating that they encrypt data without collecting or sharing data, (B) apps changing their practice from not collecting/sharing data to collecting/sharing data, and (C) apps changing their practices from collecting/sharing data to not collecting/sharing data.

For category (A) we used the following template:
\begin{enumerate}
\item[] \textit{We hope this email finds you well. We are researchers at <LAB\_NAME> and have been using your app, <APP\_NAME>, in our recent studies. We have noticed that in the data safety section of your app, it states that you encrypt data. However, we have also noticed that your app does not collect or share data.}

\textit{We are reaching out to ask if you could clarify this for us. We are trying to better understand the data safety section implemented in your app. We appreciate any information you can provide.}

\textit{Thank you for your time and we look forward to your response.}
\end{enumerate}
\section{Comparing Privacy Policies with Privacy Labels}
In this section, we provide details about how to obtain privacy practices present in the policies. We first note that \textit{Data Category} and \textit{Purpose} are the lowest levels in the taxonomy such that there are classifiers for each data category and purpose. So, extracting lower-level practices from policies is straightforward. Furthermore, we note that we have added classifiers for \textit{Encryption in Transit} and \textit{Data Deletion Option} separately. Additionally, we have two high-level classifiers, namely \textit{First-Party-Collection} and \textit{Third-party-collection} that capture segments that refer to data collection by first parties and data sharing to third parties, respectively.

To extract the remaining high-level practices, we follow the mapping shown in  \cref{tab:policy_to_label}. Specifically, as shown in \cref{tab:policy_to_label} we use multiple data category classifiers in conjunction with high-level classifier (First-party-collection-share/Third-party-collection-share) to indicate if the policy mentions data collection or data sharing. This result can be directly compared with the categories present in Google's Data Safety Section.

To match the policy to the Apple Privacy Label, we use an additional classifier: \textit{Identifiability}. This represents if the data being collected is anonymous or not. If the data is anonymous then we equate it to the \textit{Data Not Linked to You} label else the \textit{Data Linked to You} label. For example, To obtain whether a policy is collecting \textit{Location} under \textit{Data Linked To You}, we check whether there are any segments where the lower level \textit{Data Category} classifier tags the segments to contain \textit{Location} information. Then we check  whether these segments also have either \textit{First-party-collection} or \textit{Third-party-collection-share} tags in combination with \textit{Identifiability-identifiable} tag. 
\begin{figure}[ht]
    \centering
    \includegraphics[scale=0.17]{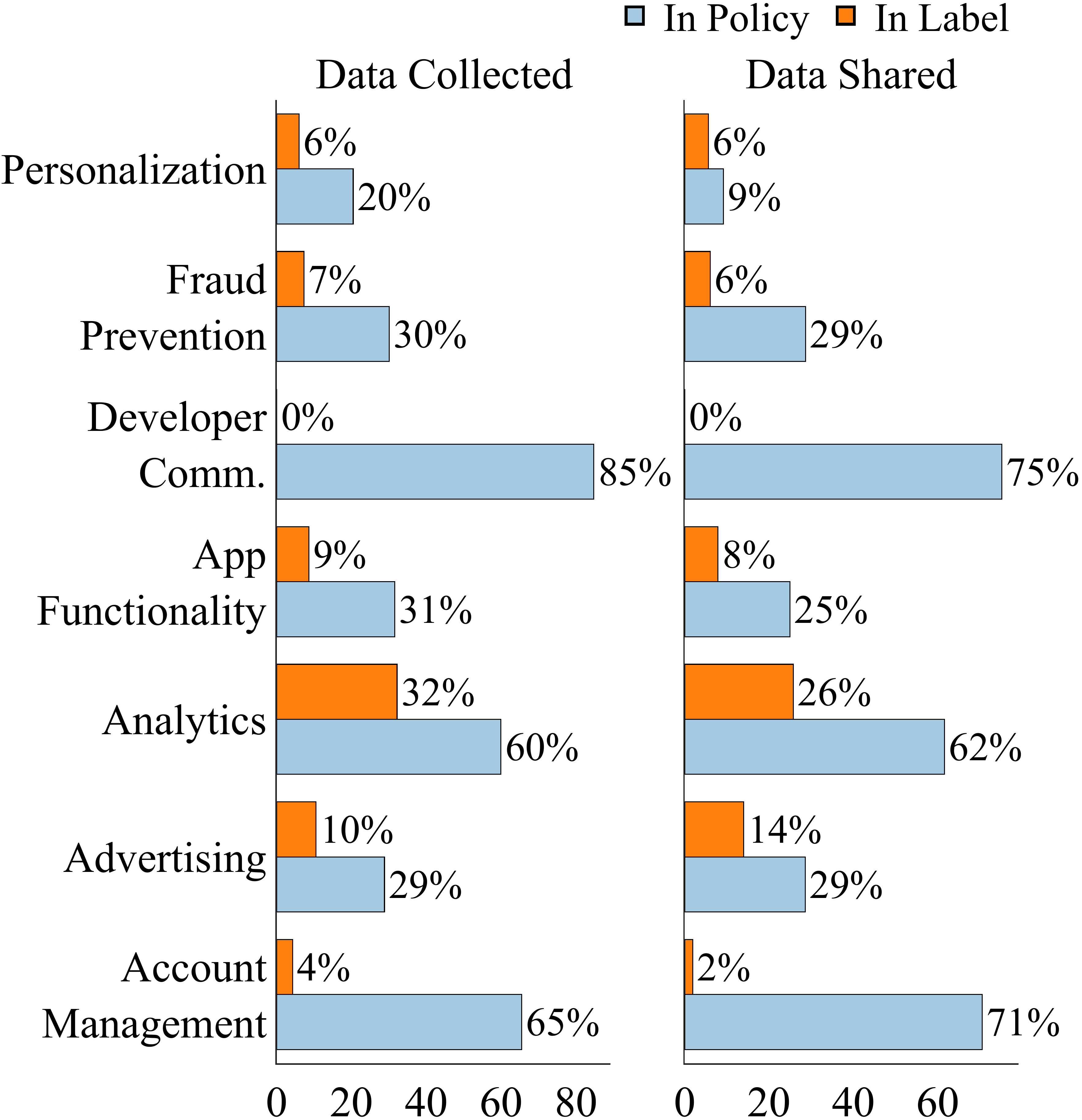}
    \caption{Inconsistencies between privacy policies and DSS, purpose based}
    \label{fig:purpose_inconst_google}
\end{figure}

\section{Mapping from DSS to APL}
\label{appendix:mapping}

In \cref{table:common_map1} and \cref{table:common_map2} we show how we convert the datatypes of DSS to those of APL. We do these conversions based on the definitions provided by Google and Apple respectively. 

\begin{figure}[ht]
    \centering
    \includegraphics[scale=0.17]{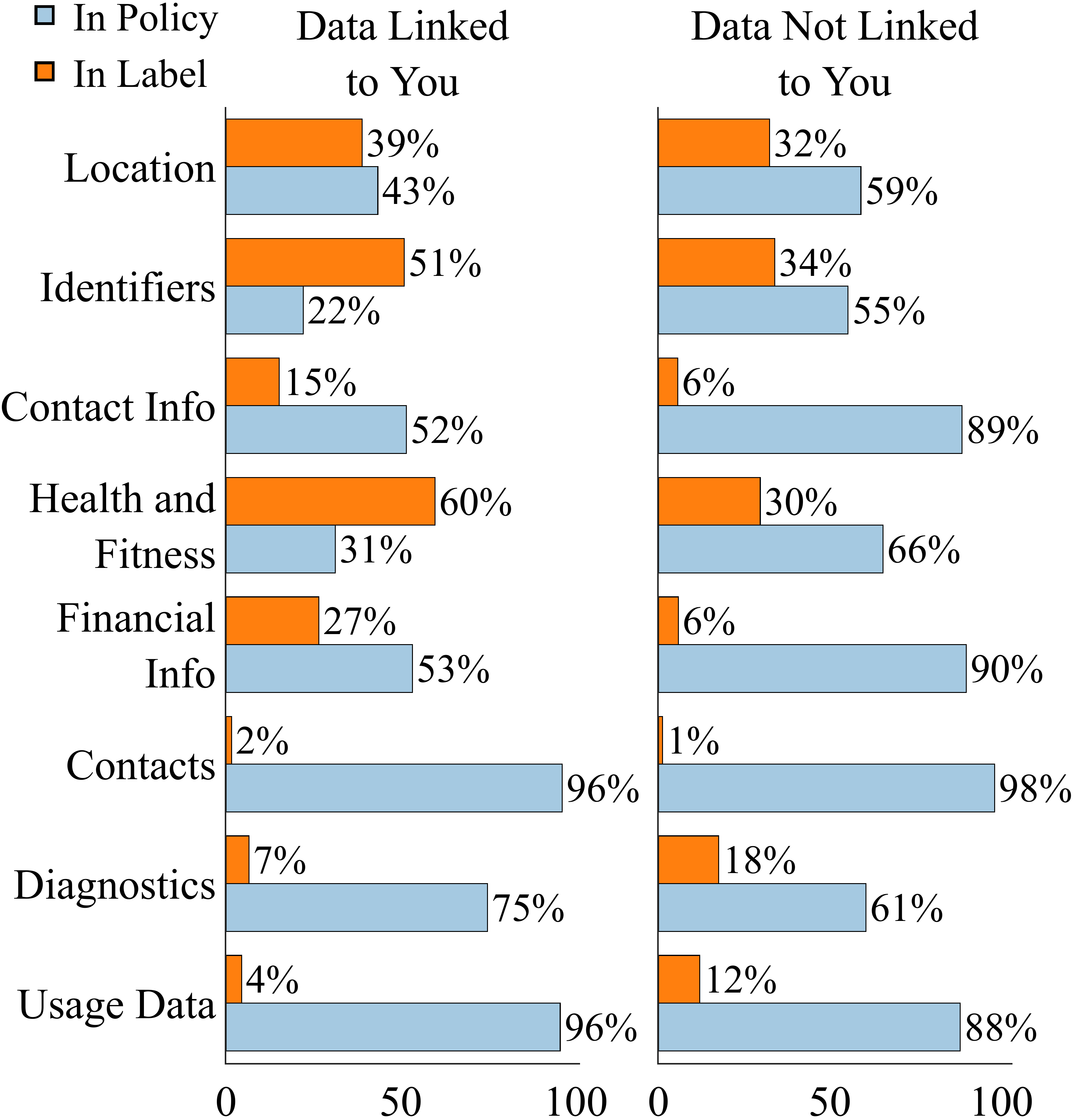}
    \caption{Inconsistencies between privacy policies and APL, datatypes based}
    \label{fig:datatype_inconst_apple}
\end{figure}
% \clearpage
\begin{figure}[ht]
    \centering
    \includegraphics[scale=0.17]{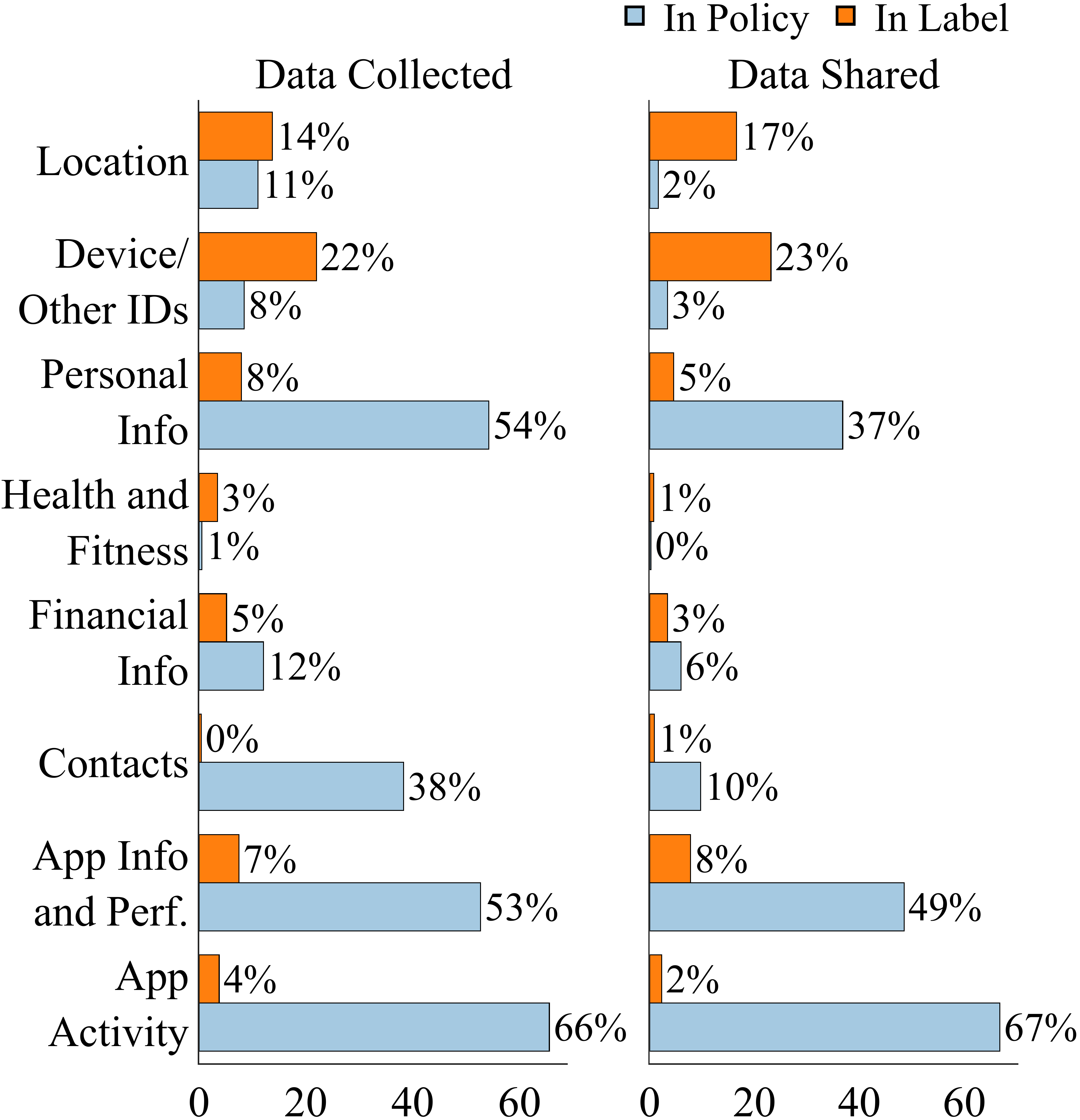}
    \caption{Inconsistencies between privacy policies and DSS, datatypes based}
    \label{fig:datatype_inconst_google}
\end{figure}
% \begin{figure}[htbp]
%   \centering
%   \includegraphics[width=\columnwidth]{figures/good_pdfs/data_safety_card_trends2_cropped.pdf}
%   \caption{Data Safety Trends}
%   \label{fig:trend_with_time}
% \end{figure}

\begin{figure}[p]
  \centering
  \includegraphics[scale=0.17]{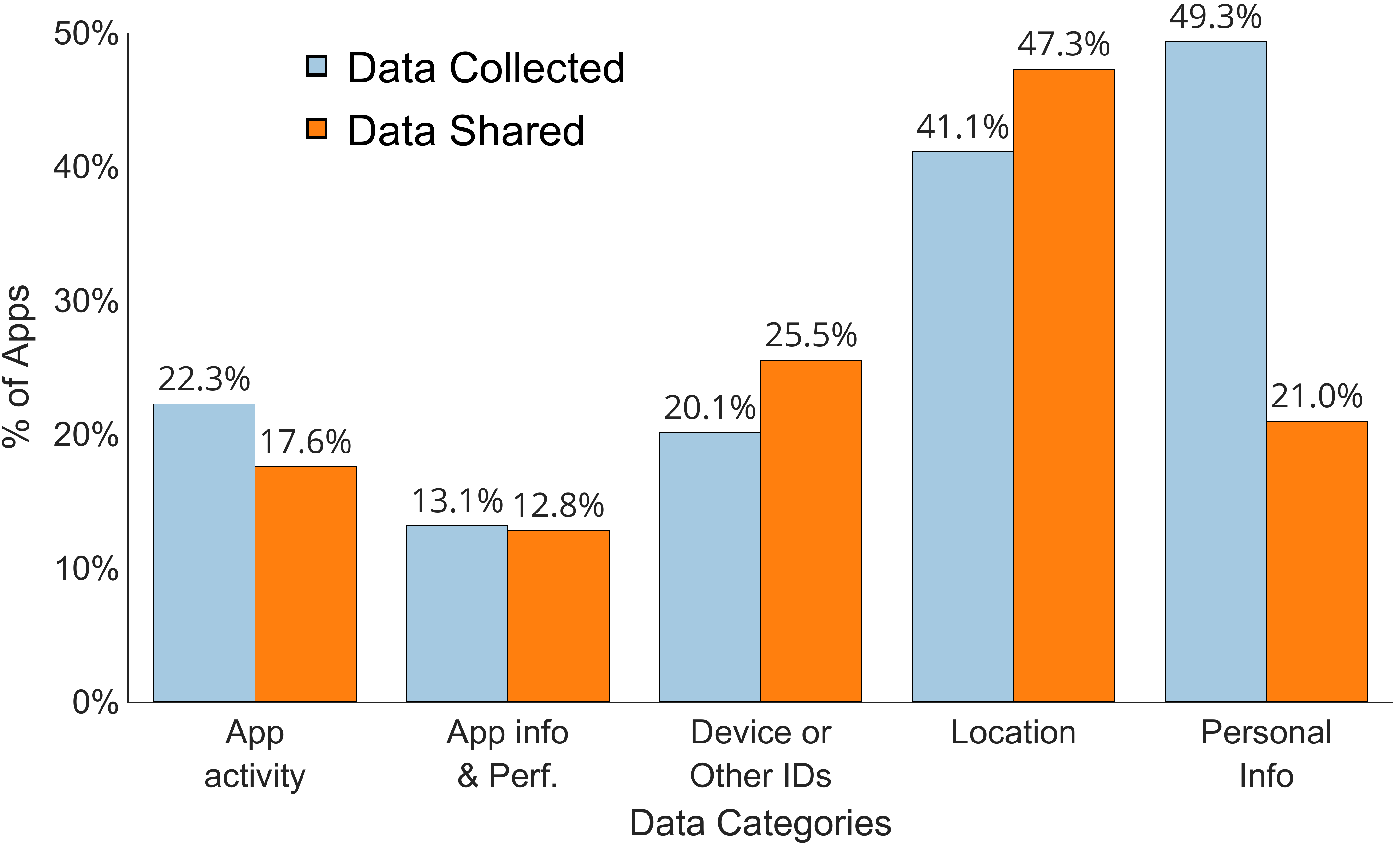}
  \caption{Plot showing what data category is collected or shared in Google Data Safety cards}
  \label{fig:google_datatype_distribution}
\end{figure}

\begin{figure}[ht]
  \centering
  \includegraphics[scale=0.8]{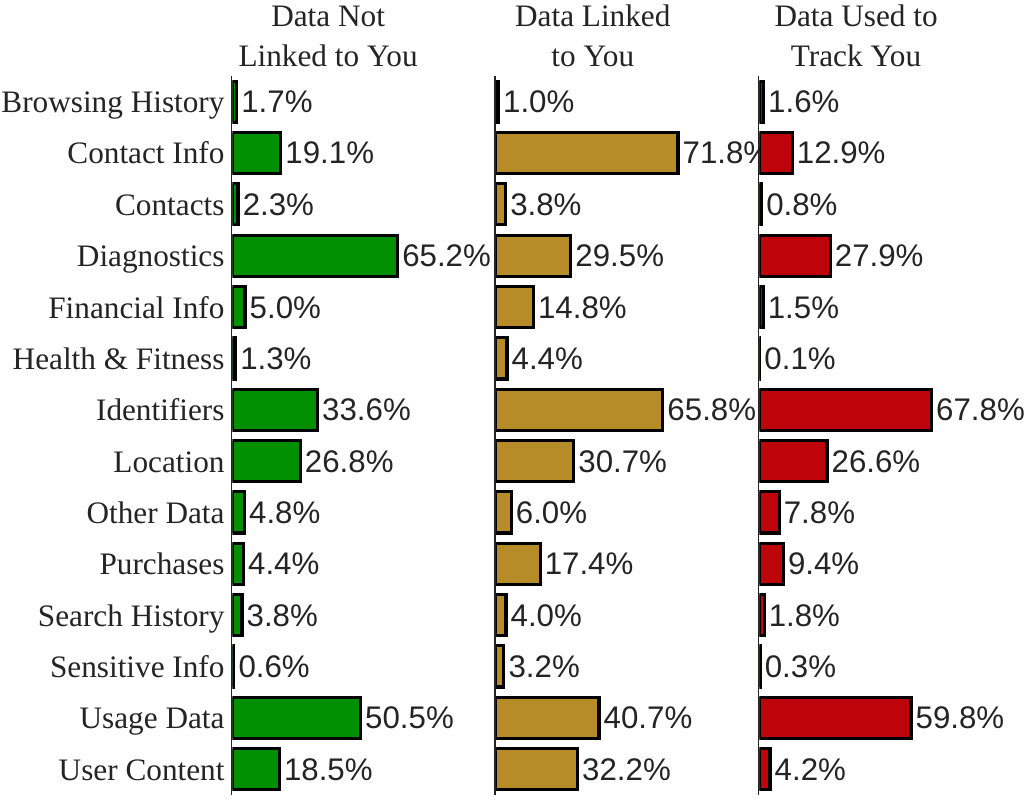}
  \caption{Plot showing the data categories used for high-level privacy practices in APL. }
  \label{fig:methodology}
\end{figure}

\begin{figure}[ht]
  \centering
  \includegraphics[width=\columnwidth]{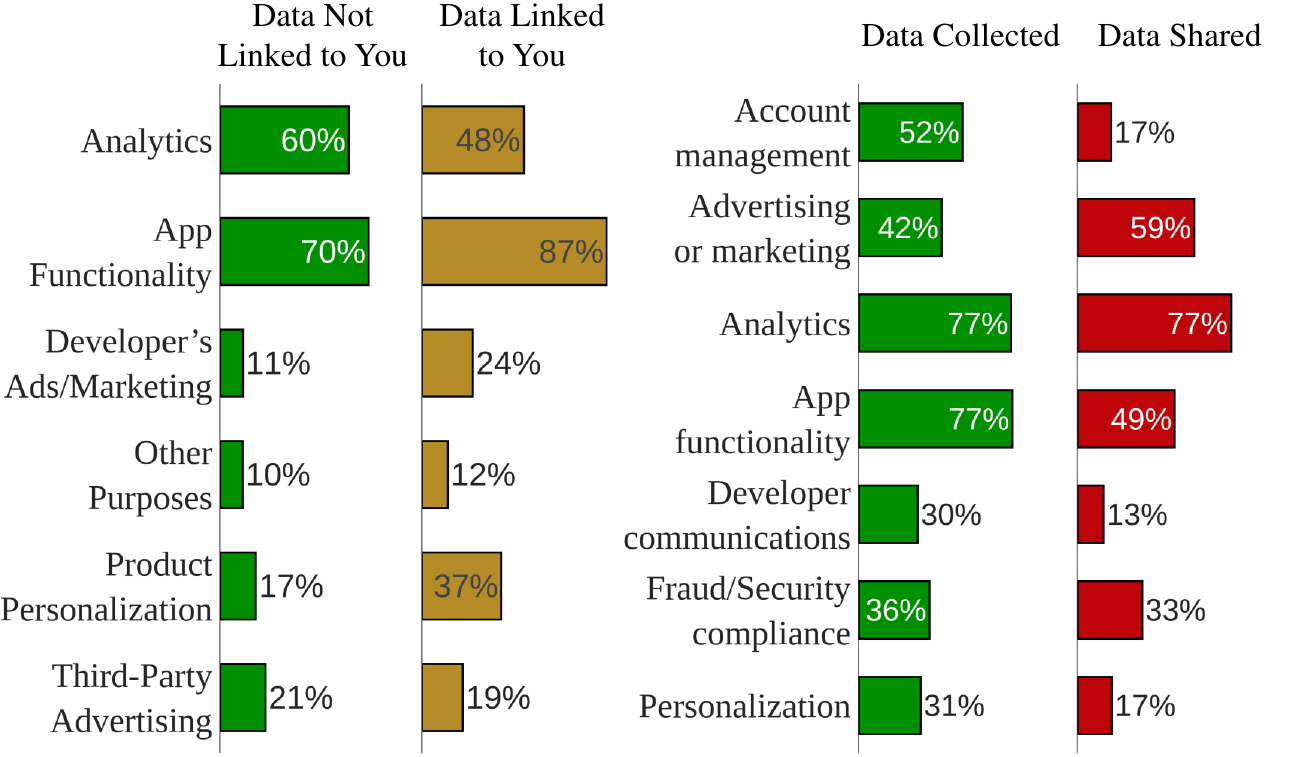}
  \caption{Plot showing the distribution of purpose for high-level privacy practices in APL}
  \label{fig:purpose_methodology}
\end{figure}

% \begin{figure*}
%   \centering
%   \includegraphics[width=2\columnwidth]{figures/good_pdfs/purpose_datatype_paircount_google_cropped.pdf}
%   \caption{Plot showing what data category is collected or shared in Google Data Safety cards}
%   \label{fig:google_datatype_purpose_distribution}
% \end{figure*}

\begin{table}[htbp]
\footnotesize
    \centering
\begin{tabularx}{\columnwidth}{|p{3.3cm}|>{\centering\arraybackslash}X|>{\centering\arraybackslash}X|>{\centering\arraybackslash}X|>{\centering\arraybackslash}X|}
\hline
\textbf{Data Category Classifiers}  & \textbf{High Level Classifier}  & \textbf{Google Data Safety Section} & \textbf{Classifier} & \textbf{Apple Privacy Labels} \\ \hline
\multirow{2}{3.3cm}{App Activity, App Info and Performance, Sensitive Info, Location, Health and Fitness, ...} & \multirow{2}{3cm}{First-party-collection-share} & \multirow{2}{3cm}{Data Collection}  & Identifiability (Identifiable) & Data Linked to You \\ \cline{4-5} 
 & & & Identifiability (Anonymous) & Data Not Linked to You \\ \hline
\multirow{2}{3.3cm}{App Activity, App Info and Performance, Sensitive Info, Location, Health and Fitness, ...} & \multirow{2}{3cm}{Third-party-collection-share} & \multirow{2}{3cm}{Data Sharing} & Identifiability (Identifiable) & Data Linked to You \\ \cline{4-5} 
& & & Identifiability (Anonymous) & Data Not Linked to You        \\ \hline
\end{tabularx}
\caption{This table shows the bottoms-up approach to get the high-level classification for Google's Data Safety Section and Apple Privacy Labels}
\label{tab:policy_to_label}
\end{table}

\begin{table}[htbp]
\centering
\begin{tabular}{lccc}
\toprule
\textbf{Category} & \textbf{CNN ~\cite{o2015introduction}} & \textbf{BERT ~\cite{devlin-etal-2019-bert}} & \textbf{Here}\\
\midrule
\rowcolor{aliceblue}First-party-collection-share  & 82 & 91 & 98 \\
Third-party-sharing-collection & 82 & 90 & 96 \\
\rowcolor{aliceblue}Identifiability & 77 & 91 & 97\\
Does-does-not & 86 & 93 & 96 \\
\rowcolor{aliceblue}Encryption-in-transit  & N/A & N/A & 99 \\
Data Deletion Option  & N/A & N/A & 91 \\
\rowcolor{aliceblue} (DC) App Activity & N/A & N/A & 93 \\
(DC) App Info and Performance & N/A & N/A & 93 \\
\rowcolor{aliceblue} (DC) Sensitive Info  & N/A & N/A & 97 \\
(DC) Location  & N/A & N/A & 99 \\
\rowcolor{aliceblue} (DC) Health and Fitness  & N/A & N/A & 97 \\
(DC) Device or Other ID  & N/A & N/A & 94 \\
\rowcolor{aliceblue} (DC) Photos and Videos  & N/A & N/A & 96 \\
(DC) Web Browsing  & N/A & N/A & 96 \\
\rowcolor{aliceblue} (DC) Contacts  & N/A & N/A & 87 \\
(DC) Calendar  & N/A & N/A & 94 \\
\rowcolor{aliceblue} (Purp) Account Management  & N/A & N/A & 92 \\
(Purp) Developer Communication  & N/A & N/A & 93 \\
\rowcolor{aliceblue} (Purp) Personalization & N/A & N/A & 98 \\
\bottomrule
\end{tabular}
\caption{Classifiers's performance on the test set. Training was done on one GPU with early stopping.}
\label{tab:privacy_label_classifier_stats}
\end{table}

\begin{table}[htbp]
\footnotesize
\centering
\begin{tabular}{>{\arraybackslash}l>{\arraybackslash}l>{\arraybackslash}l}
\toprule
\multicolumn{1}{c}{\textbf{Google Purposes}} & $\rightarrow$      & \multicolumn{1}{c}{\textbf{Apple Purposes}} \\ \midrule
Advertising or Marketing  &$\rightarrow$  & Advertising or Marketing  \\ \midrule
\rowcolor{aliceblue}Analytics   & $\rightarrow$ & Analytics\\ \midrule
App Functionality   & $\rightarrow$ &App Functionality \\ \midrule
\rowcolor{aliceblue}\begin{tabular}[l]{@{}l@{}}Fraud prevention, Security, \\ and Compliance\end{tabular} & $\rightarrow$ & App Functionality\\ \midrule
Personalization  & $\rightarrow$ & Personalization \\ \midrule
\rowcolor{aliceblue}Account Management       & $\rightarrow$ & N/A  \\  \midrule
Developer Communication  & $\rightarrow$ & N/A  \\
\bottomrule
\end{tabular}
\caption{Table showing the common mapping from Data Safety Card to Apple Privacy Label}
\label{table:common_map1}
\end{table}

\begin{table}[p]
    \footnotesize
    \centering
    \begin{tabular}{>{\arraybackslash}l>{\arraybackslash}l>{\arraybackslash}l}
    \toprule
    \multicolumn{1}{c}{\textbf{Google DataType}} & $\rightarrow$ & \multicolumn{1}{c}{\textbf{Apple DataType}} \\ \midrule
    Approximate Location & $\rightarrow$ & Coarse Location\\ \midrule
\rowcolor{aliceblue}Precise Location & $\rightarrow$ & Precise Location\\ \midrule
Name & $\rightarrow$ & Name\\ \midrule
\rowcolor{aliceblue}Email Address & $\rightarrow$ & Email Address\\ \midrule
Address & $\rightarrow$ & Physical Address\\ \midrule
\rowcolor{aliceblue}Phone Number & $\rightarrow$ & Phone Number\\ \midrule
Race And Ethnicity & $\rightarrow$ & Sensitive Info\\ \midrule
\rowcolor{aliceblue}Political Or Religious Belief & $\rightarrow$ & Sensitive Info\\ \midrule
Sexual Orientation & $\rightarrow$ & Sensitive Info\\ \midrule
\rowcolor{aliceblue}User Ids & $\rightarrow$ & User Id\\ \midrule
User Payment Info & $\rightarrow$ & Payment Info\\ \midrule
\rowcolor{aliceblue}Credit Score & $\rightarrow$ & Credit Info\\ \midrule
Other Financial Info & $\rightarrow$ & Other Financial Info\\ \midrule
\rowcolor{aliceblue}Purchase History & $\rightarrow$ & Purchase History\\ \midrule
Health Info & $\rightarrow$ & Health\\ \midrule
\rowcolor{aliceblue}Fitness Info & $\rightarrow$ & Fitness\\ \midrule
Emails & $\rightarrow$ & Emails Or Text Messages\\ \midrule
\rowcolor{aliceblue}Sms Or Mms & $\rightarrow$ & Emails Or Text Messages\\ \midrule
Other In-App Messages & $\rightarrow$ & N/A\\ \midrule
\rowcolor{aliceblue}Photos & $\rightarrow$ & Photos Or Videos\\ \midrule
Videos & $\rightarrow$ & Photos Or Videos\\ \midrule
\rowcolor{aliceblue}Voice Or Sound Recordings & $\rightarrow$ & Audio Data\\ \midrule
Music Files & $\rightarrow$ & N/A\\ \midrule
\rowcolor{aliceblue}Other Audio Files & $\rightarrow$ & N/A\\ \midrule
Files And Docs & $\rightarrow$ & N/A\\ \midrule
\rowcolor{aliceblue}Calendar & $\rightarrow$ & N/A\\ \midrule
Contacts & $\rightarrow$ & Contacts\\ \midrule
\rowcolor{aliceblue}App Interactions & $\rightarrow$ & Product Interaction\\ \midrule
Other User-Generated Content & $\rightarrow$ & Other User Content\\ \midrule
\rowcolor{aliceblue}In-App Search History & $\rightarrow$ & Search History\\ \midrule
Other Actions & $\rightarrow$ & N/A\\ \midrule
\rowcolor{aliceblue}Web Browsing History & $\rightarrow$ & Browsing History\\ \midrule
Crash Logs & $\rightarrow$ & Crash Data\\ \midrule
\rowcolor{aliceblue}Diagnostics & $\rightarrow$ & Performance Data\\ \midrule
Other App Performance Data & $\rightarrow$ & Other Diagnostic Data\\ \midrule
\rowcolor{aliceblue}Device Or Other Ids & $\rightarrow$ & Device Id\\ \midrule
Other Info & $\rightarrow$ & N/A\\ \midrule
    \bottomrule
    \end{tabular}
    % \caption{Caption}
    % \label{tab:my_label}    
    \caption{Table showing the common mapping from Data Safety Card to Apple Privacy Label}
    \label{table:common_map2}
\end{table}

\end{document}